\documentclass[aps, prd, 10pt, superscriptaddress, preprintnumbers, nofootinbib, twocolumn]{revtex4-2}
\usepackage[english]{babel}
\usepackage{bm}
\usepackage{newtxtext}
\usepackage{ulem}
\usepackage{hyperref}
\usepackage{nameref}
\hypersetup{
    colorlinks=true,
    linkcolor=blue,
    filecolor=magenta,      
    urlcolor=blue,
    citecolor=blue,
    pdftitle={Sharelatex Example},
    pdfpagemode=FullScreen
    }
\usepackage{listings}
\usepackage{slashed}
\usepackage{xcolor}
\usepackage{color}
\usepackage{appendix}
\usepackage{mathtools}
\usepackage{epsfig}
\usepackage{dcolumn}
\usepackage{textcomp}
\usepackage{appendix} 
\usepackage{multirow}
\usepackage{soul}
\usepackage{subfigure}
\usepackage{amsmath,latexsym}
\usepackage{amsmath,amssymb}
\usepackage{graphicx}
\usepackage{hyperref} 

\begin{document} 
\bibliographystyle{apsrev4-2}
\title{Quantum cosmological perturbations in bouncing models with mimetic dark matter}

\author{Idaiane L. Machado}
\email[Corresponding author: ]{idaiane.machado@hotmail.com}
\affiliation{PPGCosmo, CCE - Universidade Federal do Esp\'{\i}rito Santo, \\ Avenida Fernando Ferrari 514, 29075-910, Vitória, Esp\'{\i}rito Santo, Brazil}

\author{D\^eivid R. da Silva}
\email{deivid.rodrigo.ds@gmail.com}
\affiliation{COSMO---Centro Brasileiro de Pesquisas F\'{\i}sicas, \\ Xavier Sigaud, 150, Urca, 22290-180, Rio de Janeiro, Brazil}

\author{Nelson Pinto-Neto}
\email{nelsonpn@cbpf.br}
\affiliation{COSMO---Centro Brasileiro de Pesquisas F\'{\i}sicas, \\ Xavier Sigaud, 150, Urca, 22290-180, Rio de Janeiro, Brazil}

\begin{abstract}

We calculate the power spectrum of cosmological perturbations originated from quantum vacuum fluctuations in bouncing scenarios proposed in Ref.~\cite{chamseddine2014cosmology} in the framework of mimetic cosmology. We show that all physically relevant models produce scale-invariant spectral indices and amplitudes compatible with observations, provided that the bounce occurs at length scales $t_0$ inside the physically reasonable interval $10^5 l_p < t_0 < 10^9 l_p$. We also show that by slightly modifying the scalar field potential proposed in Ref.~\cite{chamseddine2014cosmology}, we can obtain the observed red-tilted spectral index, with the same amplitude constraints. The underlying theory contains instability problems yet to be solved. However, the methods we constructed and the results we obtained may be sufficiently general to be applied to the infrared limit of some yet to be constructed healthy mimetic cosmology, providing reasonable bouncing cosmological models without the need for any background quantum effect.

\vspace{0.2cm}

\end{abstract}


\maketitle
\flushbottom


\section{\label{Introduction} Introduction }

The Standard Cosmological Model (SCM) assumes the existence of a dark sector that constitutes approximately $96\%$ of all constituents of the Universe. One part is a cold, almost pressureless matter component that interacts weakly-or not at all-with ordinary matter, and the other one is a cosmological constant, motivating the acronym $\Lambda$CDM model. Although the $\Lambda$CDM model exhibits remarkable agreement with astrophysical and cosmological observations (see, however, Ref.~\cite{DESI}), the nature and origin of the dark sector of the Universe remain unknown.

One proposition for the physical nature of dark matter was presented in \cite{chamseddine2013mimetic}. Through a reformulation of General Relativity by isolating its conformal degree of freedom in a covariant way, A. H. Chamseddine and V. Mukhanov showed that a dynamical degree of freedom with the properties of dark matter emerges naturally in this framework, called mimetic dark matter. In this proposal, the physical metric $g_{\mu\nu}$ is rewritten in terms of an auxiliary metric $\tilde{g}_{\mu \nu}$ and a scalar field $\phi$
\begin{equation}
        g_{\mu \nu}(\tilde{g}_{\mu \nu}, \phi) = \tilde{g}_{\mu \nu} \tilde{g}^{\alpha \beta} \partial_{\alpha} \phi \partial_{\beta} \phi \; . 
        \label{eq: modmetric}
\end{equation} 
The scalar field must satisfy the constraint 
\begin{equation}
        g^{\mu \nu} \partial_{\mu} \phi \partial_{\nu} \phi=1
        \label{eq: constraint}
\end{equation}
which ensures both the physical and auxiliary metrics are invertible. This constraint is not present in other scalar-tensor theories, and it allows us to interpret $\phi$ as a velocity potential.                                                                                                                 
The tensor \( g_{\mu\nu} \) remains invariant under conformal transformations of the auxiliary metric, that is, \( g_{\mu\nu} \rightarrow g_{\mu\nu} \) when \( \tilde{g}_{\mu\nu} \rightarrow \Omega^2 \tilde{g}_{\mu\nu} \). This invariance introduces an additional degree of freedom in the equations of motion, effectively mimicking the behavior of dark matter. Subsequently, the theory was generalized through the inclusion of a potential, enabling both dark matter and dark energy to emerge as constants of integration in the resulting equations. In this framework, these components are interpreted as consequences of the underlying space-time geometry, arising from a minimal modification of General Relativity via the addition of a non-dynamical scalar field. The potential further allows for the derivation of inflationary scenarios, quintessence models, and non-singular bouncing cosmologies~\cite{chamseddine2014cosmology}.

Mimetic matter has been and continues to be investigated in various domains of cosmology and astrophysics, including spiral galaxy rotation curves \cite{myrzakulov2016static, vagnozzi2017recovering}, structure formation \cite{farsi2022structure}, cosmological perturbations \cite{matsumoto2015cosmological}, cosmological singularities, black hole physics \cite{chamseddine2017resolving, chamseddine2017nonsingular, sheykhi2021topological, chen2018black, nashed2022black, calza2024implications}, and observational constraints on the speed of gravitational waves detected by LIGO-VIRGO in events such as GW170817–GRB170817A \cite{casalino2018mimicking, Casalino2019alive, Sharafati2021higher}. This framework has also been generalized to $f(R)$ gravity and other extensions \cite{nojiri2014mimetic, myrzakulov2015inflation, cognola2016covariant, ramo2019cosmological, Chen2021thick, Mansoori2022multi, kaczmarek2024mimetic}, also encompassing disformal transformations \cite{jirouvsek2022disforming}. 

In this paper, we focus on bounce solutions obtained in the framework of Ref.~\cite{chamseddine2014cosmology} and their cosmological perturbations. Bouncing models are alternatives to avoid the initial singularity and other problems of the SCM \cite{bounceNovello,bounceNPN1,ijjas2018bouncing}. In such models, the Universe experiences a period of contraction until it bounces and emerges in the expanding phase observed today. Depending on the model being considered, this process may occur once \cite{bounceNovello,bounceNPN1} or repeatedly in the so-called cyclic models \cite{cyclic-Ijjas,cyclic-Penrose}. It is observed that the behavior of the scale factor is highly sensitive to the free parameters introduced, yielding a plethora of cosmological scenarios, depending on their chosen values. Among the possible regimes are models with a single singularity, static universes, and single- or multiple-bouncing models. We focus on single-bouncing models. We analyze the Hubble parameter, the energy density, and the scalar curvature for these scenarios. For a particular choice of the parameters, one obtains exactly the class of models presented in Ref.~\cite{bounceNPN1} with a null equation of state parameter, necessary to obtain a scale-invariant spectrum of scalar cosmological perturbations. We then examine the impact of the mimetic field within an enlarged set of parameters on the power spectrum, and we find that the resulting spectral index can also be made scale-invariant with the observed amplitude. We also show that a slight modification of the scalar field potential yields solutions identical to those presented in Ref.~\cite{bounceNPN1}, with a slightly negative equation of state parameter. As a consequence, this leads to a red-tilted spectrum of cosmological perturbations, characterized by a primordial power spectrum of the form $P(k) \propto C k^{n_s - 1}$, with spectral index $n_s < 1$ and $|n_s-1| \ll 1$. Such a red-tilt implies that larger scales (small $k$) exhibit higher relative amplitude, leading to enhanced anisotropies in the CMB at low multipoles, as observed. This result is obtained here without introducing negative pressures. Note, however, that the results we obtained must be taken with caution. The reduced scalar perturbation Hamiltonian obtained from the minimal higher-derivative mimetic action is not bounded from below \cite{Firouzjahi_2017, IJJAS2016132}. This indicates that the minimal model contains ghost instabilities and should not be regarded as a complete ultraviolet theory, but rather as a low-energy effective description, requiring a UV completion or additional stabilizing operators, whose effects on the infrared limit are yet to be understood.

This paper is organized as follows. In Section \ref{Mimetic matter} we summarize mimetic theory and its extension to cosmology; in Section \ref{Mimetic matter and non-singular bounce} we present a variety of background bouncing models emerging in this framework. Afterwards, in Section \ref{Perturbations}, we compute the power spectrum of cosmological perturbations in such models, showing that a class of them yields the observed primordial power spectrum. A summary of our conclusions and outlook is presented in the final section. We adopt the space-time signature as (+, -, -, -).


\section{\label{Mimetic matter}Mimetic matter review} 

The action of mimetic gravity can be written as \cite{chamseddine2014cosmology}
\begin{equation}
 \begin{split} 
        S  =& \frac{1}{l_p^2}\int d^{4} x \sqrt{-g} \bigg[- \frac{1}{2} R (g_{\mu \nu}) + \\ & \mathbf{\lambda} (g^{\mu \nu} \partial_{\mu} \phi \partial_{\nu}\phi - 1) - V (\phi)+\mathcal{L}_{m}(g_{\mu \nu}, ...) \bigg],
            \label{eq: action}
 \end{split}
\end{equation}
where $l_p \equiv 8\pi \mathcal{G}$ is the Planck length (we adopt natural units with $\hbar = c = 1$), $\mathcal{G}$ is Newton's gravitational constant, $g$ is the determinant of the physical metric, and $R$ is the Ricci scalar. The $\lambda$ is a Lagrange multiplier that enforces the mimetic constraint, while $\mathcal{L}_m$ denotes the Lagrangian for matter, which depends on the metric and the fields of matter. The function $V(\phi)$ is an arbitrary potential for the scalar field. This class of action was first presented in three publications \cite{lim2010dust, capozziello2010dark, gao2011cosmological}, see \cite{sebastiani2017mimetic} for a review.

By varying the action~\eqref{eq: action} with respect to the Lagrange multiplier $\lambda$, we recover the constraint~\eqref{eq: constraint}. Varying with respect to the physical metric $g_{\mu \nu}$ and the mimetic field $\phi$ yields two equations of motion respectively
\begin{equation}
        G_{\mu \nu}-2\lambda \partial_{\mu} \phi \partial_{\nu} \phi-g_{\mu \nu} V(\phi)=T_{\mu \nu},
        \label{eq: motion1}
\end{equation}
\begin{equation}
        \nabla^\mu(\lambda \partial_\mu \phi)  + \frac{dV}{d\phi} = 0,
        \label{eq: motion2}
\end{equation}
where $G_{\mu\nu}$ and $T_{\mu\nu}$ are the Einstein and the energy-momentum tensors of the physical metric and the usual matter, respectively. Note that we have absorbed $8\pi \mathcal{G}$ in the definition of $T_{\mu\nu}$ (and in the forthcoming $\tilde{T}_{\mu \nu}$), $8\pi \mathcal{G}$ $T_{\mu\nu}\rightarrow T_{\mu\nu}$, thus $\rm{dim} [T_{\mu\nu}] =$ $(\rm{length})^{-2}$ in what follows. 

Taking the trace of equation (\ref{eq: motion1}) allows us to express the Lagrange multiplier as 
\begin{equation}
        \lambda = \frac{1}{2} (G - T -4V),
        \label{eq: lagrangem}
\end{equation}
where $G = g^{\mu\nu} G_{\mu\nu}$ and  $T = g^{\mu\nu} T_{\mu\nu}$ are their respective traces. Replacing Eq.~(\ref{eq: lagrangem}) into Eqs. ~(\ref{eq: motion1}), we get $G_{\mu \nu} = T_{\mu \nu} + \tilde{T}_{\mu \nu},$ where $\tilde{T}_{\mu \nu} \equiv (G - T -4V)\partial_{\mu} \phi \partial_{\nu} \phi+g_{\mu \nu} V(\phi)$. The comparison between $\tilde{T}_{\mu \nu}$ with the energy-momentum tensor of a perfect fluid form $T_{\mu \nu} = (\varepsilon + p) u_{\mu} u_{\nu} - pg_{\mu \nu}$ leads to
\begin{equation}
        \tilde{p} = -V,
        \label{eq: pressure}
\end{equation}
\begin{equation}
        \tilde{\varepsilon} = (G - T -3V),
        \label{eq: energy density}
\end{equation}
\begin{equation}
       u_\mu = \nabla_\mu \phi ,
        \nonumber
\end{equation}
as its pressure, energy density, and fluid $4-$velocity (as said above, $\phi$ playing the role of a velocity field thanks to Eq.~\eqref{eq: constraint}), respectively. Thus, equation (\ref{eq: motion1}), with (\ref{eq: lagrangem}), is equivalent to Einstein's equations with an additional fluid characterized by pressure $\tilde{p}$ and energy density $\tilde{\varepsilon}$. The scalar field corresponds to the potential velocity, while the constraint (\ref{eq: constraint}) is the normalization condition for the 4-velocities. 

Assuming the background to be homogeneous, isotropic, and spatially flat, equations (\ref{eq: motion1}) and (\ref{eq: constraint}) imply that the scalar field is only a function of time, $\phi = \pm t + A$, where $A$ is a constant of integration. Without loss of generality, we will identify the scalar field with time. Consequently, as the pressure and energy density in equations (\ref{eq: pressure}) and (\ref{eq: energy density}) are time-dependent only, using the combination of them allows us to rewrite equation (\ref{eq: motion2}) as
\begin{equation}
        \frac{1}{a^3} \frac{d}{dt}(a^3 (\tilde{\varepsilon}-V)) = - \frac{dV}{dt},
        \label{eq: motion22}
\end{equation}
where $a$ is the scale factor, yielding 
\begin{equation}
        \tilde{\varepsilon} = \frac{C}{a^3} +V - \frac{1}{a^3} \int a^3 \dot{V} dt = \frac{C}{a^3}+  \frac{3}{a^3} \int a^2 V da,
        \label{eq: energy}
\end{equation}
where $C$ is an integration constant associated with the mimetic dark matter component. For a non-vanishing potential $V$, there is an additional contribution to the energy density of the mimetic matter, which is similar to a generalized cosmological constant added to the Lagrangian and identical to it when $V=$ constant.

The time-time component of equation (\ref{eq: motion1}) plus (\ref{eq: lagrangem}) is the Friedmann equation
\begin{equation}
        H^2 = \frac{8\pi G}{3} \tilde{\varepsilon} = \frac{8\pi G}{a^3} \int a^2 V da \; ,
        \label{eq: f}
\end{equation}
where $H \equiv \dot{a}/a$ is the Hubble function. Note that the constant $C$ was not set to zero; rather, it was absorbed into the integration constant arising from $\int V a^2 da$ in the equation above. Taking its derivative with respect to time yields
\begin{equation}
        2 \dot{H}+ 3H^2 = V.
        \label{eq: friedmann}
\end{equation}
Introducing the new variable $y = a^{\frac{3}{2}}$ enables us to rewrite equation (\ref{eq: friedmann}) in the simpler form,
\begin{equation}
        \ddot{y} - \frac{3}{4} V(t) y = 0,
        \label{eq: diff}
\end{equation}
which is a linear differential equation that allows us to find homogeneous and isotropic cosmological solutions.


\section{\label{Mimetic matter and non-singular bounce} Mimetic matter and non-singular bounce}

Following \cite{chamseddine2014cosmology}, we consider the action in the presence of mimetic potential $V(\phi)$ plus the addition of the last extra term,
\begin{equation}
 \begin{split} 
        S =& \frac{1}{l_p^2}\int d^{4} x \sqrt{-g} \bigg[- \frac{1}{2} R (g_{\mu \nu})  + \\ & \mathbf{\lambda} (g^{\mu \nu} \partial_{\mu} \phi \partial_{\nu}\phi - 1) -  V (\phi) + \frac{1}{2} \gamma (\Box \phi)^2 \bigg],
        \label{eq: modaction}
 \end{split}
\end{equation}
where $\gamma$ is a dimensionless constant, and $\Box = g^{\mu\nu} \nabla_\mu \nabla_\nu$.
The addition of this term does not affect the homogeneous background cosmological solutions, preserves the number of degrees of freedom, and plays a crucial role in perturbation theory by introducing a non-null sound velocity for the perturbations, as we will see. This extra term was studied theoretically and phenomenologically in Refs. \cite{capela2015modified, ramazanov2016living, babichev2017gravitational} where constraints on the constant $\gamma$ have been obtained.

Varying the action \eqref{eq: modaction} with respect to the 
physical metric, and defining $\Gamma \equiv \Box \phi$, yields (see Ref.~\cite{chamseddine2014cosmology}) 
\begin{align}
        G^\mu_\nu &= \left[ V + \gamma \left( \phi_{,\alpha} \Gamma^{,\alpha} + \frac{1}{2} \Gamma^2 \right) \right] \delta^\mu_\nu \nonumber + \\
        &\quad + 2 \lambda \phi_{,\nu} \phi^{,\mu} - \gamma \left( \phi_{,\nu} \Gamma^{,\mu} + \Gamma_{,\nu} \phi^{,\mu} \right) = 
        \tilde{T}^\mu_\nu.
        \label{15}
\end{align}
As before, the general solution for $\phi$ in the homogeneous and isotropic case reads $\phi = t + A$, consequently $\Gamma = \Box \phi = \ddot{\phi} + 3H\dot{\phi} = 3H $. The 0-0 component of the Einstein equation and the i-j components read, respectively, 
\begin{equation}
        H^2 = \frac{1}{3}V + \gamma \bigg( \frac{3}{2} H^2 - \dot{H}\bigg) + \frac{2}{3} \lambda,
        \label{eq: mod0-0c}
\end{equation}
\begin{equation}
        2 \dot{H}+ 3H^2 = \frac{2}{2 - 3\gamma} V.
        \label{eq: modfriedmann}
\end{equation}
Note that there is now a new constant multiplying the potential $V$, see equation (\ref{eq: friedmann}). As a result, the cosmological solutions derived in Ref.~\cite{chamseddine2014cosmology} remain unchanged up to this numerical factor of order unity. In particular, depending on the form of $V$, one can have $\dot H > 0$, allowing for the occurrence of a bounce. Using again $y = a^{\frac{3}{2}}$, we obtain the following differential equation,
\begin{equation}
        \ddot{y} - \frac{3}{2(2-3\gamma)} V(t) y = 0.
        \label{eq: moddiff}
\end{equation}

Different cosmological scenarios can be obtained from the differential equations (\ref{eq: diff}) and (\ref{eq: moddiff}) depending on the potential to be worked on. Let us consider the behavior of mimetic matter in the case of bouncing situations without singularities, as presented in Ref. ~\cite{chamseddine2014cosmology}. For this, consider the potential,  
\begin{equation}
        V (\phi) = \frac{2}{3} \frac{(2 - 3\gamma)\alpha }{(\phi_0^2 + \phi^2)^2} \nonumber = \frac{2}{3} \frac{(2 - 3\gamma)\alpha }{(t_0^2 + t^2)^2},
        \label{eq: tpotential}
\end{equation}
where $\phi_0 = t_0$ is a constant. Therefore, we can rewrite the potential as,
\begin{equation}
        V t_0^2 = \frac{2}{3} \frac{(2 - 3\gamma)\alpha_0 }{(1 + \tau^2)^2},
        \label{eq: taupotential}
\end{equation}
where $\alpha_0 = \alpha/t_0^2$ and $\tau = t/t_0$. Note that this potential is always positive definite and finite as long as $\alpha_0 > 0$ and  $0<\gamma<2/3$. In this case, the differential equation \eqref{eq: moddiff} becomes
\begin{equation}
        \frac{d^2y}{d\tau^2}  - \frac{\alpha_0}{(1 + \tau^2)^2} y = 0.
        \label{eq: modtaudiff0}
\end{equation}
The general solution reads, 
\begin{equation}
\begin{aligned}
        a(\tau) ={} & \, a_b (\tau^2 + 1)^{1/3} \\
        & \times \left[ \cos(\beta \arctan{\tau}) + A \sin(\beta \arctan{\tau}) \right]^{2/3},
        \label{eq: a(tau)}
\end{aligned}
\end{equation}
where $\beta = \sqrt{1 - \alpha_0}$, therefore $\alpha_0 \leq 1$, and $A, a_b$ are constants. 

\begin{figure}[htb!]
    \centering
    \includegraphics[width = \linewidth]{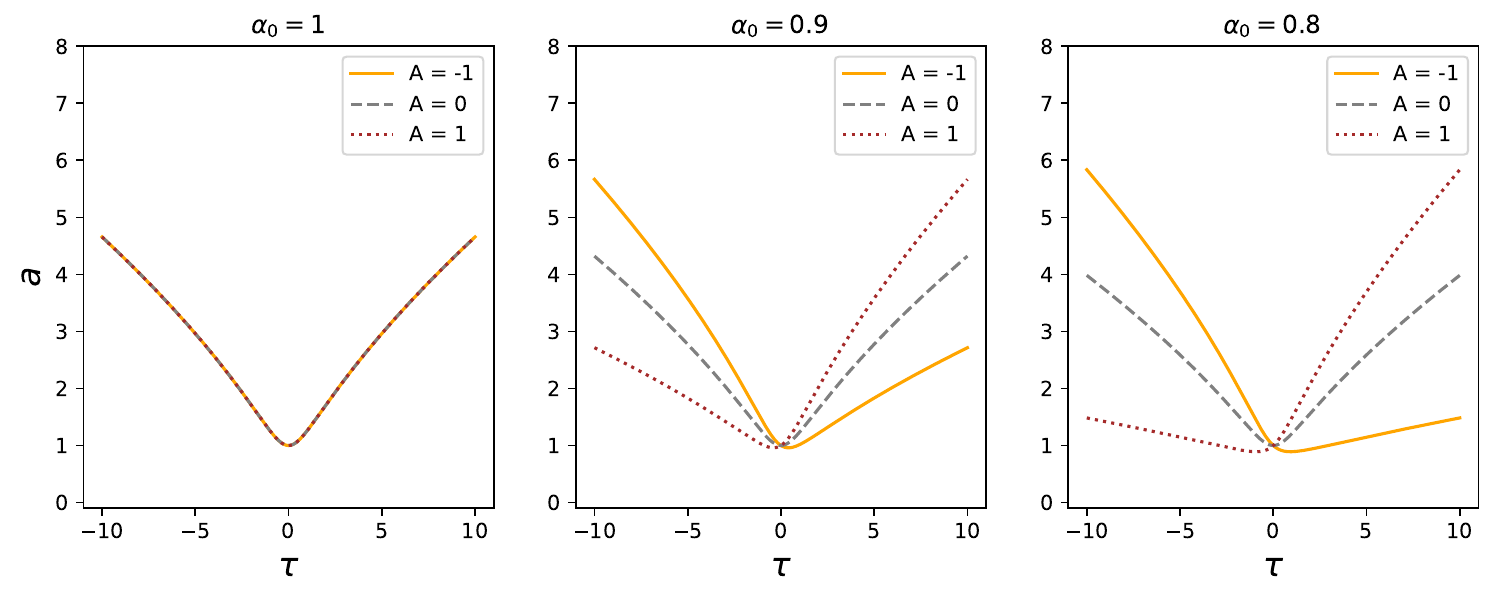}
    \caption{Evolution of the factor $a$ with the dimensionless time parameter $\tau$, considering $0.8 \leq \alpha_0 \leq 1$ and $-1 \leq A \leq 1$. All these cases are bouncing scenarios without singularities. Note that for $\alpha_0\neq 1$ and $A\neq 0$ the models are asymmetric with respect to the bounce, otherwise they are symmetric.}
    \label{tau_x_a_for_the_relevant_cases}
\end{figure}

The model considered here is rich, offering many possible cosmological scenarios that can be classified into three cases: (I) single-bounce solutions without singularities, which can be time-symmetric or time-asymmetric with respect to the bounce; see Figure \ref{tau_x_a_for_the_relevant_cases}; (II) big-bang-like models with or without a big-crunch; and (III) a static universe. This diversity in the dynamics of the scale factor is governed by the parameters $\alpha_0$ and $A$, as described in Eq.~(\ref{eq: a(tau)}). 

The singularities appear whenever
\begin{equation}
        |A|>A_{\rm min}\equiv |(\tan(\sqrt{1-\alpha_0}\;\pi/2)^{-1}|,
        \label{eq: aamin}
\end{equation}
obtained by imposing that the scale factor presented in Eq.~(\ref{eq: a(tau)}) vanishes at finite time. Conversely, the condition $|A|\le A_{\rm min}$ ensures that $a(\tau)>0$ for all times, preventing the trigonometric factor in Eq.~(\ref{eq: a(tau)}) from reaching zero. As a consequence, curvature invariants as well as the effective energy density remain finite throughout the evolution, thus avoiding big-bang or big-crunch singularities. This condition therefore selects regular bouncing solutions characterized by $H=0$ and $\dot H>0$ at the bounce.
Furthermore, $A_{\rm min}$ increases with $\alpha_0$ and diverges in the limit $\alpha_0 \to 1$, corresponding to the complete absence of singularities. As we are interested in non-singular bouncing models, we restrict, without loss of generality, the parameter space to $0.75 \leq \alpha_0 \leq 1$ and $-1 \leq A \leq 1$. It is worth noting that the special case $\alpha_0 = 1$ yields a scale factor that is insensitive to the parameter $A$ and coincides with the solutions obtained in \cite{bounceNPN1}.

The asymptotic time limits of Eq.~\eqref{eq: a(tau)} read,
\begin{equation}
\begin{aligned}
        \lim_{\tau \to \pm \infty} a(\tau) ={}& \; a_b 
        \left[ \cos\left( \frac{\beta \pi}{2} \right) 
        \pm A \sin\left( \frac{\beta \pi}{2} \right) \right] |\tau|^{2/3} \\
        & + \mathcal{O}(|\tau|^{-1/3}) + \ldots \;,
\end{aligned}
\label{eq: asympt1}
\end{equation}
showing that the bounce is asymmetric unless $A=0$ or $\beta=0$. 
As is well known, the leading time-dependence $a(\tau) \approx |\tau|^{2/3}$ is typical of a pressureless fluid domination, like dark matter, showing that the mimetic field can indeed represent dark matter when $\tau >> 1$, far from the bounce phase. Using the Friedmann equation in these asymptotic limits, one gets
\begin{equation}
        \lim_{\tau\to \pm \infty}a(\tau) \approx a_0
        \left(\frac{9}{4} \Omega_{0m}^{\pm}\frac{t_0^2}{R_{H0}^2}\right)^{1/3}|\tau|^{2/3},
        \label{eq: asympt2}
\end{equation}
where $\Omega_{0m}^{\pm}\equiv \rho_{0m}^{\pm}/\rho_{\rm cr}$, $\rho_{0m}^{\pm}$ are the energy densities of dark matter at the asymptotic expanding and contracting phases, respectively, and $\rho_{\rm cr}, R_{H0}$ are the critical density and Hubble radius today, respectively. From these equations, we get
\begin{equation}
        \Omega_{0m}^{-} = \left(\frac{\cos(\beta \pi/2) - A \sin(\beta \pi/2)}{\cos(\beta \pi/2) + A \sin(\beta \pi/2)}\right)^2 \Omega_{0m}^{+} .
        \label{eq: density}
\end{equation}
It should be noted that $\Omega_{0m}^{-}$ may be smaller than or greater than $\Omega_{0m}^{+}$ depending on the sign of $A$. Note also that for $\beta \ll 1$, we obtain
\begin{equation}
        \tau_{\rm bounce} \approx -A\beta,
        \label{eq: taubounce}
\end{equation}
\begin{equation}
        a(\tau_{\rm bounce})\approx a_b (1-A^2 \beta^2/3),
        \label{eq: abounce}
\end{equation}
\begin{equation}
        \Omega_{0m}^{+}\approx\frac{4}{9 x_b^3} \frac{R_{H0}^2}{t_0^2}(1+\pi A\beta).
        \label{eq: omegapbouce}
\end{equation}
Here, $\tau_{\mathrm{bounce}}$ is the time of the bounce, defined by $H(\tau_{\mathrm{bounce}}) = 0$, and $x_b = a_0 / a_b$.  
The last expression above relates the observable quantities $\Omega_{0m}^{+}$ and $R_{H0}$ to the theoretical parameters $t_0$, $a_b$, $\beta$, and $A$.

Figure \ref{tau_x_a_for_the_relevant_cases} shows these bounce solutions. For $\alpha_0 =1$, we recover the symmetrical bounce solution presented in Ref.~\cite{bounceNPN1}. For $0.75 \leq\alpha_0 < 1$, we get asymmetric bounce solutions. 
As the asymptotic behavior of the scale factor is $a\propto \tau^{2/3}$ in all cases, the universe contracts from the far past dominated by a highly diluted cold matter until it reaches a bounce around $\tau_{\rm bounce}$. The concentrated matter has a compression limit, and the scalar curvature reaches a maximum value when the scale factor reaches a minimum. This situation is unstable, and the field launches the universe into an expanding phase. After a certain time interval, the scalar field returns to mimic cold dark matter.\footnote{Of course, at some specific moment, radiation, baryons, and dark energy should play a role, maybe created by the strong gravitational field around the bounce, but we will focus our attention to the contracting phase, the bounce, and shortly after it.}

\begin{figure}[htb!]
    \centering
    \includegraphics[width = \linewidth]{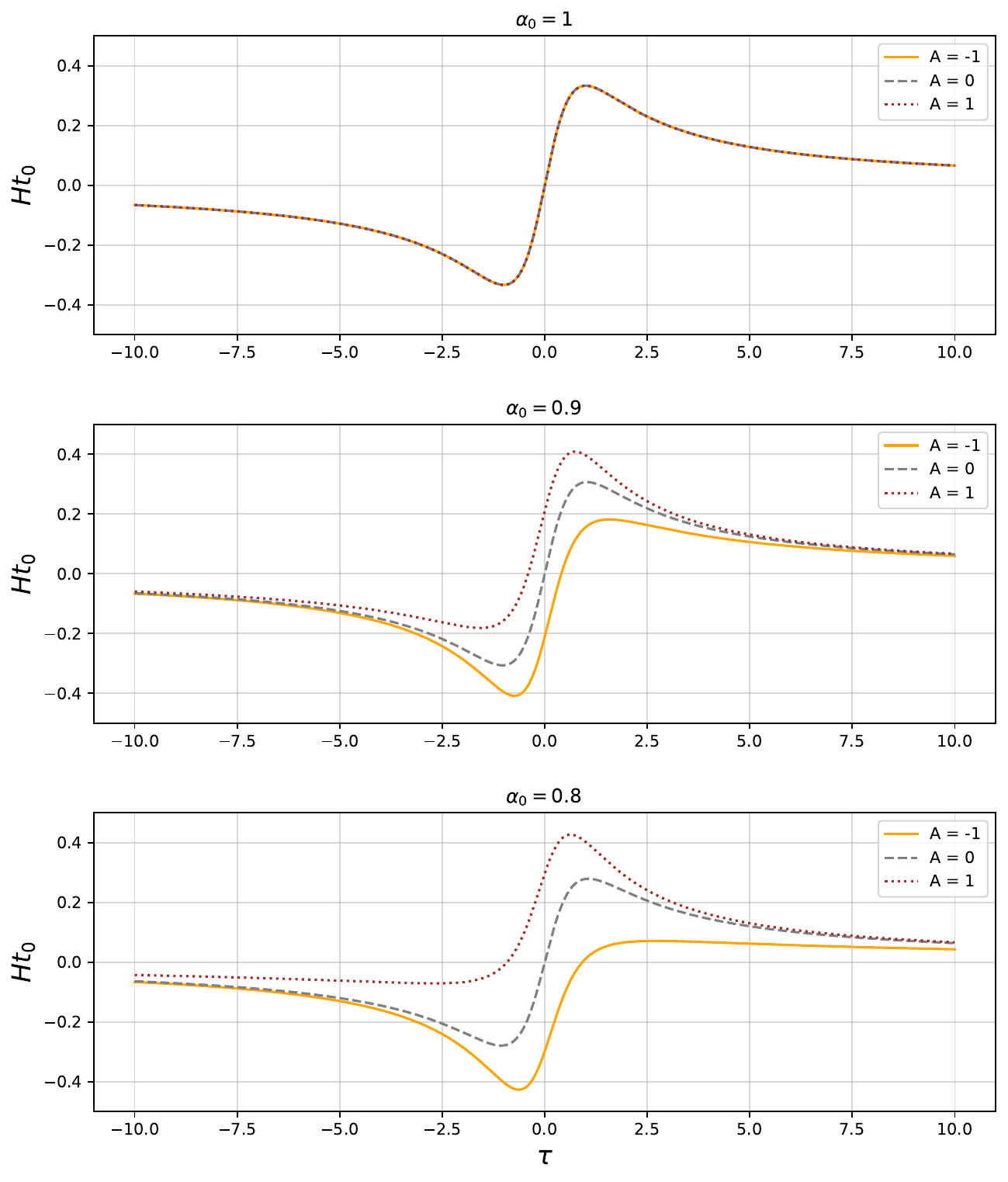}
    \caption{The Hubble parameter $H$ evolution for models with a unique bounce event around $\tau \approx 0$.}
    \label{fig: tau_x_H}
\end{figure}

\begin{figure}[htb!]
    \centering
    \includegraphics[width = \linewidth]{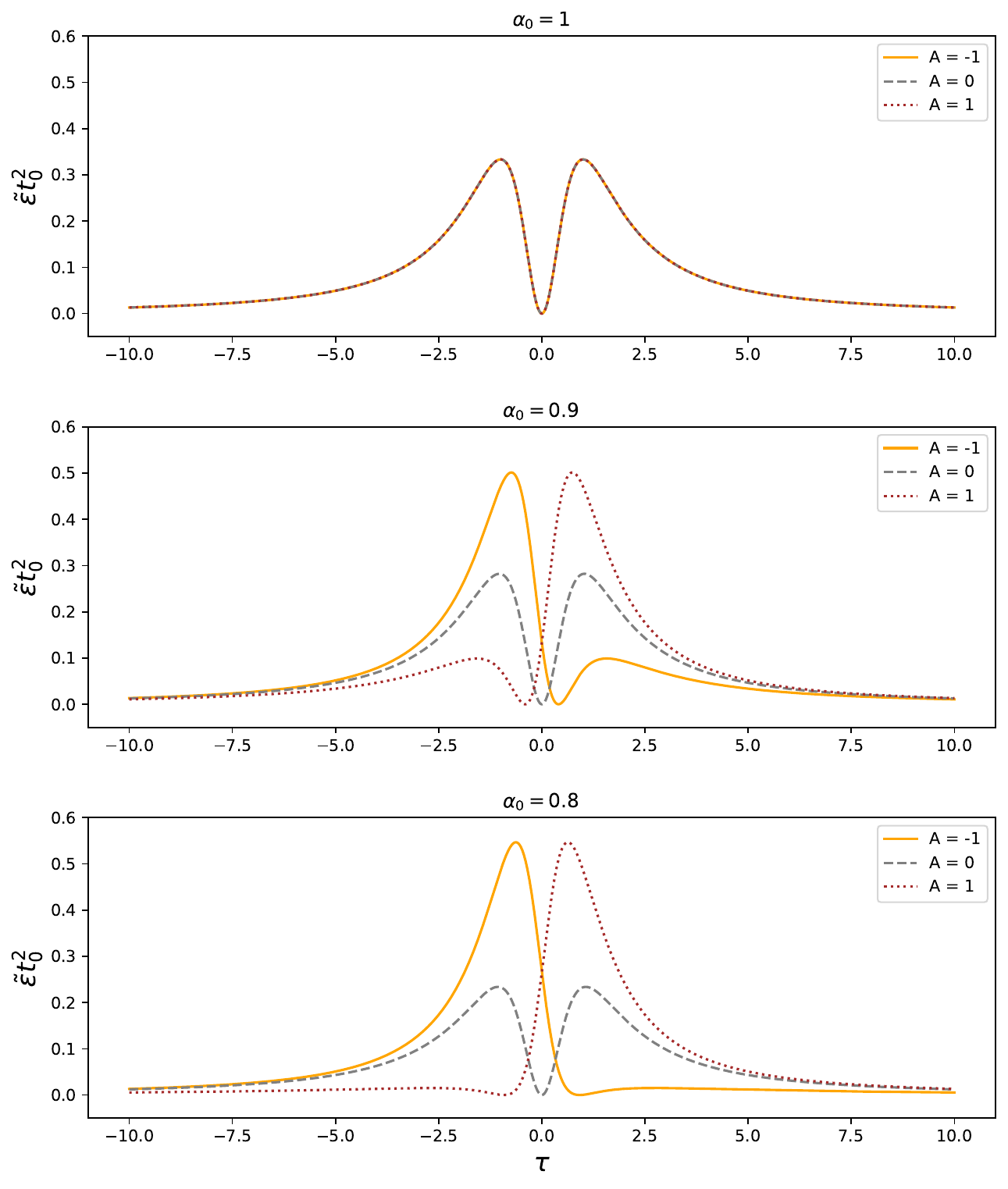}
    \caption{The energy density $\tilde{\epsilon} = H^2$ evolution for models with a unique bounce event around $\tau \approx 0$.}
    \label{fig: tau_x_energy_density}
\end{figure}

In Figures \ref{fig: tau_x_H} and \ref{fig: tau_x_energy_density}, we plot the Hubble parameter and the energy density of the scalar field, calculated using the Friedmann equation, as functions of $\tau$ in this parameter domain. The cases with $\alpha_0 = 1$ or $A = 0$ present the symmetric behavior discussed before: the contraction occurs at the same rate as the expansion phase (with opposite signs), and the energy density presents the same behavior in both phases. The asymmetries emerge when $0.75<\alpha_0<1 $ and $A\neq 0$, where one can see that the scalar field presents bigger energy densities in the expanding (contracting) phase for models with $A$ positive (negative), in accordance with Eq.~\eqref{eq: density}, signaling creation (destruction) of the scalar field energy around the bounce. In all cases, the energy density goes to zero in the asymptotic limits of $\tau$. Note also that the bouncing instants of time one can read in these figures are in accordance with Eqs.~\eqref{eq: taubounce} and ~\eqref{eq: abounce}.

\begin{figure}[htb!]
    \centering
    \includegraphics[width = \linewidth]{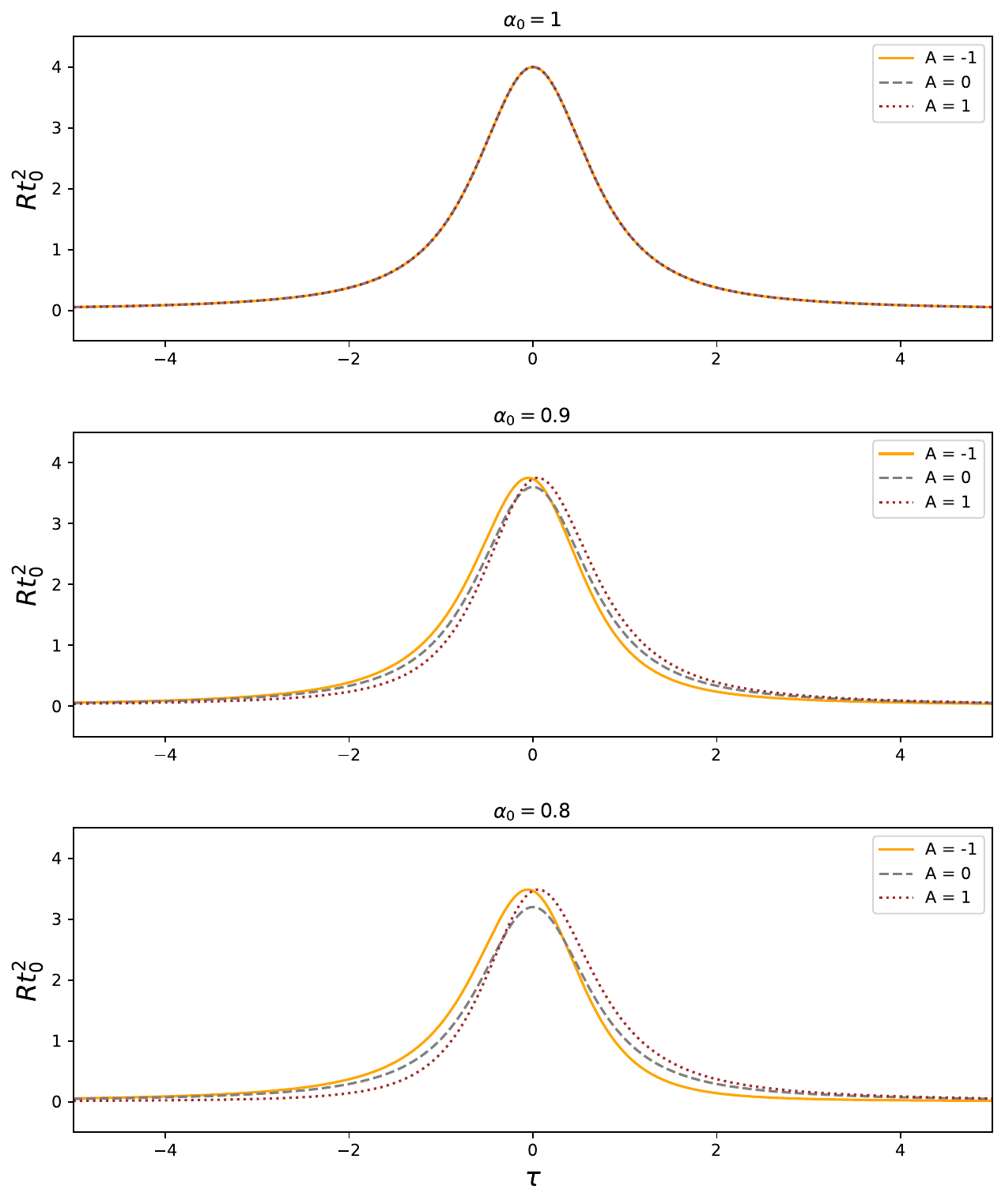}
    \caption{The Ricci scalar $R$ evolution for models with a unique bounce event around $\tau \approx 0$.}
    \label{fig: tau_x_Ricci_Scalar}
\end{figure}

The scalar curvature is calculated by the relation
\begin{equation}
        R = \frac{6}{t_0^2}\left[ \frac{1}{a}\frac{d^2a}{d\tau^2} + \left( \frac{1}{a} \frac{da}{d\tau}\right)^2 \right],
\end{equation}
and as the reader can see in Fig. \ref{fig: tau_x_Ricci_Scalar}, in all cases of interest, the asymptotic behavior in the far past or future is $R\rightarrow 0$, i.e., the spacetime curvature approaches flat space. Again, the scalar curvature presents a symmetric or asymmetric evolution around the bounce, depending on the choice of parameters, as discussed earlier. 

One can see from Figs.~(3,4) that neither the energy density nor the scalar curvature gets close to the Planck scale along the whole evolution of the model. Hence, we expect that quantum gravity effects turn out to be irrelevant for it, as they are expected to be important only at these scales. Indeed, taking $t_0\gg t_p$, where $t_p$ is the Planck time, from Fig.~(3) $R_{\rm{max}} \approx 4/t_0^2 \ll 1/t_{\rm{Planck}}^2 \approx R_{\rm{Planck}}$. The same reasoning can be used using Figs.~(4) getting that $\epsilon_{\rm{max}} \ll \epsilon_{\rm{Planck}}$.
Note that one should also impose that $t_0\ll t_N$, where $t_N$ is the characteristic time-scale of nucleosynthesis, in order to demand that the energy scale of the bounce should be much bigger than the nucleosynthesis energy scale, in order not to affect it.

%

\section{Perturbations}\label{Perturbations}

In this section, we will perturb the bouncing models presented in the previous section in order to calculate the power spectrum of scalar cosmological perturbations in such mimetic field scenarios and compare with some observational data. For that, we consider the perturbations in the metric in the Newtonian gauge:
\begin{equation}
        ds^2 = [1 + 2\psi(t, \mathbf{x})]dt^2 - [1 - 2\psi(t, \mathbf{x})]a^2(t)d\mathbf{x}^2,
        \label{eq:metric with perturbations}
\end{equation}
where $\psi(t, \mathbf{x})$ is the Newtonian potential. We also consider a first-order perturbation of the scalar field, i.e.,
\begin{equation}
        \phi(t, \mathbf{x}) = \bar{\phi}(t) + \delta \phi(t, \mathbf{x}).
\end{equation}

The constraint $g^{\mu \nu}\partial_\mu \phi \partial_\nu \phi = 1$ produces a zeroth-order equation $\dot{\bar{\phi}}~^2 = 1$ and a first-order equation $\psi = \dot{\bar{\phi}} \delta \dot{\phi}$. One solution to the zeroth-order equation is $\bar{\phi} = t$, which generates the following connection between the Newtonian potential and the perturbation of the scalar field:
\begin{equation}
        \psi = \delta \dot{\phi}.
        \label{eq: delta and psi relation}
\end{equation}
Therefore, to obtain the power spectrum of $\psi$, which is the one confronted with observations, we only have to calculate the power spectrum of $\delta\dot\phi$. 
Manipulating the first-order terms of the Einstein equations yields the following differential equation for the scalar field perturbations \cite{chamseddine2014cosmology}:
\begin{equation}
    \delta \ddot{\phi} + H\delta\dot{\phi} - \frac{c_s^2}{a^2}\nabla^2\delta \phi + \dot{H}\delta \phi = 0,
    \label{delta phi}
\end{equation}
where $c_s^2=\frac{\gamma}{2-3\gamma}.$ Note that if $\gamma =0$, the sound velocity of the perturbations would be null, which is discarded by observations. Furthermore, $\gamma$ should be small and positive if one wants the mimetic scalar field to indeed represent dark matter.
In conformal time $\rm{d}t = a \rm{d}\eta$, where $a$ is the scale factor, and for the Fourier modes $\delta\phi_k$ we get,
\begin{equation}
        \delta {\phi_k}'' + \left[c_s^2k^2 + \frac{a''}{a} - 2 \left(\frac{a'}{a}\right)^2\right] \delta \phi_k = 0,
        \label{mode-phi}
\end{equation}
where a prime designates derivative with respect to conformal time.
This is a set of second-order differential equations for each $k$ with general solution $C_1(k) U_{k1}(\eta) + C_2(k) U_{k2}(\eta)$, where $U_{k1}(\eta), U_{k2}(\eta)$ are two independent solutions of Eq.~\eqref{mode-phi}, and $C_1(k), C_2(k)$ are the two integration constants for each $k$, hence being arbitrary functions of $k$. To obtain the power spectrum, one needs to specify $C_1(k), C_2(k)$. The usual approach in bouncing models, similar to what is done in inflation, is to assume that in the far past of such models, which is an almost empty and flat universe, as it can be seen from figures (2-4), where the Hubble function, energy density, and scalar curvature goes to zero in this limit, the inhomogeneities are dissipated and they can exist only as quantum vacuum (adiabatic) fluctuations; see Ref.~\cite{bounceNPN1}. These quantum fluctuations are then enhanced due to the growth of the gravitational field while the universe contracts, becoming the classical seeds of the structures we see today in the Universe \cite{Ward1, Ward2}. Consequently, it is necessary to quantize the perturbations, which requires identifying the appropriate canonical variable for quantization. Note that, as said above, the parameters of the model were chosen in such a way that the background does not need to be quantized; hence, we quantize perturbations in a classical background. As the perturbations are fields, we are facing quantum field theory in a curved background, and the vacuum must be chosen according to well-known physical requirements; see Refs.~\cite{Mukhanov2005BookOfCosmology,Sandro}.


\subsection{Quantization and analytical results}
\label{subsection: Quantization and analytical results}
Expanding the action \eqref{eq: modaction} to the second order in perturbations and isolating the kinetic term of the scalar field in the action yields
\begin{equation}
     \frac{1}{2l_p^2}\int d\eta d^3\mathbf{x} \frac{\gamma}{c_s^2} \partial_i \delta \phi^\prime \partial_j \delta \phi^\prime \delta^{ij} \supset S,
\end{equation}
where $S$ is the action. In order to put the kinetic term in the canonical form for quantization, $(v^\prime)^2/2$, one defines the quantum canonical variable as
\begin{equation}
    v(x) = \frac{\sqrt{\gamma}}{c_s}\partial_i\delta\phi(x) .
    \label{eq: v definition}
\end{equation}
It follows that $v(x)$ is dimensionless, as $\delta\phi(x)$ has dimensions of length (time), and it has a "mass" $m=1/l_p^2$.

From the mode version of Eq.~\eqref{eq: v definition},
\begin{equation}
        v_k\propto\frac{\sqrt{\gamma}}{c_s}k\delta\phi_k,  
        \label{vk phik}
\end{equation} 
the mode action for $v_k$ arising from the full action \eqref{eq: modaction} reads
\begin{equation}
        \frac{1}{2l_p^2}\int d\eta [- v_k^{\prime 2} + (c_s^2 k^2 + a^2 \dot{H})v_k^2],
        \label{mode-v action}
\end{equation}
yielding the equations of motion

\begin{equation}
        v_k'' + \left[c_s^2k^2 + \frac{a''}{a} - 2 \left(\frac{a'}{a}\right)^2\right] v_k = 0,
        \label{mode-v}
\end{equation}

Quantizing $v(x)$ as usual, one gets
\begin{equation}
\label{v-operator}
\hat{v}(\eta,\mathbf{x})=\frac{1}{\sqrt{2}}\int \frac{\rm{d}^3k}{(2\pi)^{3/2}}[v_k(\eta)e^{i\mathbf{k}\cdot\mathbf{x}}\hat{a}_k+v_k^*(\eta)e^{-i\mathbf{k}\cdot\mathbf{x}}\hat{a}_k^{\dagger}].
\end{equation}

Equation \eqref{mode-v} for the modes $v_k$ appearing in Eq.~\eqref{v-operator} is a consequence of the quantization procedure of fields in time-dependent backgrounds; see Refs.~\cite{Mukhanov2005BookOfCosmology,Sandro}. Note that making $a^\prime =0$ and getting back to Minkowski spacetime, Eq.~\eqref{mode-v} yields $v_k = \exp (\pm ik\eta)$, as usual.

As the creation and annihilation operators must have dimensions of $\rm{length}^{3/2}$ ($\int\rm{d}^3k [\hat{a},\hat{a}_k^{\dagger}]=1$), we conclude that $v_k(\eta)$ must also have dimensions of $\rm{length}^{3/2}$.

As discussed above, the adiabatic vacuum initial conditions are set in the asymptotic past, where spacetime is almost flat. Because the scale factor behaves as $a \propto |\tau|^{2/3} \propto \eta^2$ in this limit, Eq.~\eqref{mode-v} becomes
\begin{equation}
        {v_k}'' + \left[c_s^2k^2 - \frac{6}{\eta^2}\right] v_k = 0,
        \label{mode-v asympt}
\end{equation}
with a general solution
\begin{equation}
         v_k(\eta) =  C_1(k) U_{k1}(\eta) + C_2(k) U_{k1}^*(\eta),
         \label{mode-v solution}
\end{equation}
where
\begin{equation}
        U_{k1}(\eta) = e^{-ic_s k\eta}\left[1-\frac{3}{(c_s k\eta)^2}(ic_s k\eta+1)\right].
        \label{mode1-v solution}
\end{equation}
The adiabatic vacuum prescription \cite{Sandro} is posed at sub-Hubble scales, where $|c_s k\eta| \gg 1$ (which is asymptotically valid for any scale as $\eta\to -\infty$), and Eq.~\eqref{mode-v asympt} represents a bunch of harmonic oscillators for each $k$. In this situation, we have
\begin{equation}
        v_k(\eta) \approx \frac{\exp(-i\int \nu \rm{d}\eta)}{\sqrt{m\nu}} =  \frac{l_p}{\sqrt{c_s k}}e^{-ic_s k\eta},
        \label{mode-v vacuum}
\end{equation}
as $\nu = c_s k$ (see Eq.~\eqref{mode-v asympt}) and $m=1/lp^2$. Hence, Eq.~\eqref{mode-v vacuum} implies that $ C_1(k)=\frac{l_p}{\sqrt{c_s k}},\;\; C_2(k)=0.$ 
It is important to emphasize Eq. ~\eqref{mode-v asympt} is valid for any $|\tau|=|t|/t_0\gg 1$. As $t_0$ must be much smaller than the nucleosynthesis time scale, this interval encompasses a large interval of cosmic time, including the period when the large scales of cosmological interest for CMB measurements are super-Hubble. Accordingly, one can say that the solution
\begin{equation}
        v_k(\eta) \approx  \frac{l_p}{\sqrt{c_s k}}e^{-ic_s k\eta}\left[1-\frac{3}{(c_s k\eta)^2}(ic_s k\eta+1)\right],
        \label{mode-v vacuum2}
\end{equation}
is also valid in the super-Hubble limit, except near the bounce, where $|\tau|\gg 1$ is no longer satisfied. To pass through the bounce up to the expanding phase, we can match Eq.~\eqref{mode-v vacuum2} with the general long wavelength solution
\begin{equation}
        v_k(\eta) =  \frac{1}{a}\left(A_1 + A_2 \int a^2 \rm{d}\eta\right)+\rm{O(}k^2)+...,
        \label{mode-v vacuum3}
\end{equation}
resulting in $A_1\propto k^{-5/2}, A_2\propto k^{5/2}$.

From the above results it is clear that the spectrum of $v_k$ and $\delta\phi_k$ cannot be flat. However, the spectrum which should be compared with CMB observations for long wavelengths of scalar perturbations is the spectrum of the metric perturbation $\psi_k$, which is nearly proportional to the curvature perturbation $\zeta_k$ in the periods when the expanding phase era is dominated by a single fluid with $w=p/\rho$, and both $\psi_k$ and $\zeta_k$ are nearly constants; see Ref.~\cite{Mukhanov2005BookOfCosmology},

\begin{equation}
       \zeta = -\psi + \frac{H}{\dot{H}}(\dot{\psi} + H\psi) \approx -\frac{5+3w}{3(1+w)} \psi.
        \label{psi-zeta}
\end{equation}
Hence, we need to calculate
\begin{equation}
        \delta_{\psi}^2=\frac{k^3}{2\pi^2} |\psi_k|^2=\frac{k^3}{2\pi^2} |\delta\dot{\phi}_k|^2,
        \label{power spectrum}
\end{equation}
see Eq.~\eqref{eq: delta and psi relation}. Therefore, what is important is the spectrum of $\dot{\delta\phi}_k$, not $\delta\phi_k$, and the time derivative makes a crucial difference. In order to see this, we have to connect $\delta\dot{\phi}_k$ with $v_k$, which obeys Eq.~\eqref{mode-v}. 

In order to simplify the calculations, we will use the variable $\chi_k\equiv a \delta\phi_k$, and the resulting mode equation coming from equation \eqref{delta phi} reduces to
\begin{equation}
        \ddot{\chi}_k - H\dot{\chi}_k + \frac{c_s^2k^2}{a^2}\chi_k  = 0.
        \label{eq X}
\end{equation}
The above equation can be obtained from the simple Hamiltonian
\begin{equation}
        H_k = \frac{\Pi_k^2}{2m} + \frac{m \nu^2 \chi_k^2}{2},
        \label{eq H_k}
\end{equation}
with $m=1/a$ and $\nu=c_sk/a$, yielding the equivalent first order equations of motion
\begin{subequations}
    \begin{equation}
        \dot\chi_k = \frac{\partial H}{\partial \pi_k} = a\pi_k,
        \label{eq: 51a}
    \end{equation}
    \begin{equation}
        \dot\pi_k = - \frac{\partial H}{\partial \chi_k} = - \frac{(c_s k)^2}{a^3} \chi_k.
        \label{eq: 51b}
    \end{equation}
    \label{eq: system for chi and pi}
\end{subequations}

The long wavelength expansion of the solutions of Eqs.~\eqref{eq: system for chi and pi} reads
\begin{eqnarray}
        \chi_k(\eta) &=&  B_1\left(1 - \int a \rm{d}t\int\frac{c_s^2 k^2}{a^3}\rm{d}t_1\right)+\nonumber \\
        &&B_2 \left(\int a \rm{d}t - \int a \rm{d}t\int\frac{c_s^2 k^2}{a^3}\rm{d}t_1\int a \rm{d}t_2\right) + \nonumber\\
        &&\rm{O(}k^4)+...\;.
        \label{mode-x long}  
\end{eqnarray}
As $\chi_k\equiv a \, \delta\phi_k\propto a\,v_k/k$, see Eq.~\eqref{vk phik}, then the $k$-dependence of $B_1$ and $B_2$ must satisfy $B_1=A_1/k$ and $B_2 = A_2/k$.

Note that $\dot{\chi_k}/a$ is just the momentum defined in Eq.~\eqref{eq: 51a}, and it is nearly constant, as its time derivative in Eq.~\eqref{eq: 51b} is negligible for large wavelengths ($k<<1$). This constant is not zero, as it will be seen in the sequel, and it is always much bigger than $H \chi_k /a$. This statement was also verified numerically. Thus, during the expansion phase, we neglect the decaying term $\frac{H\chi_k}{a}$ in the relation $\dot{\delta\phi_k}= \frac{\dot\chi_k}{a} - \frac{H\chi_k}{a} \approx \frac{\dot\chi_k}{a}$. Hence, we get
\begin{eqnarray}
        &&\dot{\delta\phi_k} \approx \frac{\dot\chi_k}{a} = - B_1 \int\frac{c_s^2 k^2}{a^3}\rm{d}t \nonumber\\ && +B_2 \left(1-\int\frac{c_s^2 k^2}{a^3}\rm{d}t\int a \rm{d}t_1\right) + 
         \rm{O(}k^4)+...\;,
        \label{mode-dotx long} 
\end{eqnarray}
where the time derivative brings down the $k^2$ factor in the first term, multiplied by $B_1$.

In the expanding phase, the dominant part is the constant mode, but in bouncing models it receives a contribution from the term multiplying $B_1$ due to the long contracting phase, which exceeds the $B_2$ term by many orders of magnitude; see Ref.~\cite{Sandro0}. This implies that the $k$ dependence of the power spectrum \eqref{power spectrum} is
\begin{equation}
    \delta_{\psi}^2 = C k^3 (k^2 B_1)^2 = C k^3 (k A_1)^2,
    \label{power spectrum index}
\end{equation}
which is scale invariant as $A_1\propto k^{-5/2}$ from the quantum vacuum initial conditions, and $C$ is a $k-$independent quantity which can only be evaluated numerically.

As a final comment, note from Eq.~\eqref{power spectrum} that the power spectrum can be calculated solely from the canonical momentum conjugated to $\chi_k$, $\pi_k$,

\begin{equation}
        \delta_{\psi}^2=\frac{k^3}{2\pi^2} |\psi_k|^2=\frac{k^3}{2\pi^2} |\delta\dot{\phi}_k|^2 \approx\frac{k^3}{2\pi^2} |\frac{\dot\chi_k}{a}|^2=\frac{k^3}{2\pi^2} |\pi_k|^2,
        \label{power spectrum pi}
\end{equation}
which satisfies the simple Hamilton equations 
\eqref{eq: system for chi and pi}. This enormously simplifies the numerical calculations that will be described in the next subsection, showing the suitability of the $\chi_k$ variable.


\subsection{Numerical calculations}

To evaluate the amplitude of the spectrum, which is the $k-$ independent quantity $C$ appearing in Eq.~\eqref{power spectrum index}, we have to resort to numerical calculations. We will write the scale factor \eqref{eq: a(tau)} as $a_b D g(\tau)$ where 
\begin{equation}
        D\equiv \left[\cos(\beta \pi/2) - A \sin(\beta \pi/2) \right]^{2/3},
        \label{D}
\end{equation} 
hence $\lim_{\tau\to - \infty}g(\tau) = |\tau|^{2/3},$ and within the parameter range we chose, $D\approx\rm{O}(1)$.

We will work with the dimensionless variables $\chi_k^{\rm{num}}$ and $\pi_k^{\rm{num}}$ defined by
\begin{eqnarray}
 \chi_k &=&  \sqrt{\frac{c_s}{2\gamma}} \frac{l_p t_0^{3/2}}{\sqrt{a_b D}} \frac{\chi_k^{\rm{num}}}{\bar{k}^{3/2}}\nonumber \\
\pi_k &=&  \sqrt{\frac{c_s}{2\gamma}} \frac{l_p t_0^{1/2}}{(a_b D)^{3/2}} \frac{\pi_k^{\rm{num}}}{\bar{k}^{3/2}},
    \label{mode-xpi long}  
\end{eqnarray}
where $\bar{k}\equiv kt_0/(a_b D)$. The Hamilton equations \eqref{eq: system for chi and pi} now read,
\begin{subequations}
    \begin{equation}
        \frac{\rm{d}\chi_k^{\rm{num}}}{\rm{d}\tau} = g(\tau)\pi_k^{\rm{num}},
    \end{equation}
    \begin{equation}
        \frac{\rm{d}\pi_k^{\rm{num}}}{\rm{d}\tau} = - \frac{(c_s \bar{k})^2}{g^3(\tau)} \chi_k^{\rm{num}}
    \end{equation}
    \label{eq: system for chi and pi0}
\end{subequations}
with initial conditions given by
\begin{subequations}
    \begin{equation}
        \chi_k^{\rm{num,in}} = - \frac{1}{3(c_s \bar{k})^2},
    \end{equation}
    \begin{equation}
        \pi_k^{\rm{num,in}} = - \frac{1}{3\tau}.
    \end{equation}
    \label{eq: ini}
\end{subequations}
These initial conditions come from Eq.~\eqref{mode-v vacuum2}, and the relationship between $v_k$ and $v^{\prime}_k$ with $\chi_k$ and $\pi_k$ through Eq.~\eqref{vk phik} and $\chi_k=a \delta\phi_k$, keeping the appropriate terms of the expansion in $1/\tau$.

The system of equations \eqref{eq: system for chi and pi0} is very suitable for numerical solution. We achieve this using a Python library called SciPy \cite{scipy}. This library can only solve equations in the form $dU/d\tau = f(U, \tau)$, where $f$ is a generic function. Thus, in our case, $U = (\chi, \pi)$ and f is the two-vector provided by the left-hand side of Eq. (\ref{eq: system for chi and pi0}).

As in the expanding phase we have $\delta\dot{\phi_k}=\dot{\chi_k}/a - H\chi_k/a \approx  \dot{\chi_k}/a = \pi_k$, the power spectrum \eqref{power spectrum} reads
\begin{equation}
    \delta_{\psi}^2=\frac{k^3}{2\pi^2} |\delta\dot{\phi}_k|^2 = \frac{k^3}{2\pi^2} |\pi_k|^2.
    \label{power spectrum2}
\end{equation}
Expressing it in terms of the numerical variables, we get
\begin{equation}
    \delta_{\psi}^2 = \frac{c_s}{2\gamma} \frac{l_p^2}{t_0^2}\delta_{\psi, \rm{num}}^2,
    \label{power spectrum3}
\end{equation}
where
\begin{equation}
    \delta_{\psi, \rm{num}}^2 = \frac{|\pi_k^{\rm{num}}|^2}{2\pi^2} ,
    \label{power spectrum4}
\end{equation}

The wave numbers of CMB cosmological interest are around $1 < k_H < 10^4$, where $k_H\equiv k R_{H0}/a_0$ (scales from the Hubble radius to  \cite{aghanim2020planck}. The relation between $\bar{k}$ and $k_H$ reads
\begin{equation}
\bar{k} = \left(\frac{4}{9\Omega_{0,m}^+ D^3}\right)^{1/3}
\left(\frac{t_0}{R_{H0}}\right)^{1/3} k_H \approx 10^{-20} \left(\frac{t_0}{l_p}\right)^{1/3}k_H ,
\end{equation}
where we have used Eqs.~\eqref{eq: taubounce}, ~\eqref{eq: abounce} and ~\eqref{eq: omegapbouce}.

For a first numerical calculation we set $c_s \approx \sqrt{\gamma} = 10^{-5}$, following \cite{soundspeed}, and $t_0/lp = 10^9$, yielding $10^{-17} < \bar{k} < 10^{-13}$ to infer the amplitudes of $\delta_{\psi, \rm{num}}^2$ for different values of the parameters. Then we can adjust $t_0$ to give the observed amplitude of $\delta_{\psi}^2\approx 10^{-9}$. As we expect the power spectrum to be scale invariant, the new range of $\bar{k}$ adjusted to the new $t_0$ will not affect the amplitude; therefore, we can keep the same range as the one chosen. Note that $t_0$ must satisfy $l_p<<t_0<<l_N \approx 10^{21} l_p$. 

\begin{figure}[htb!]
    \centering
    \includegraphics[width = \linewidth]{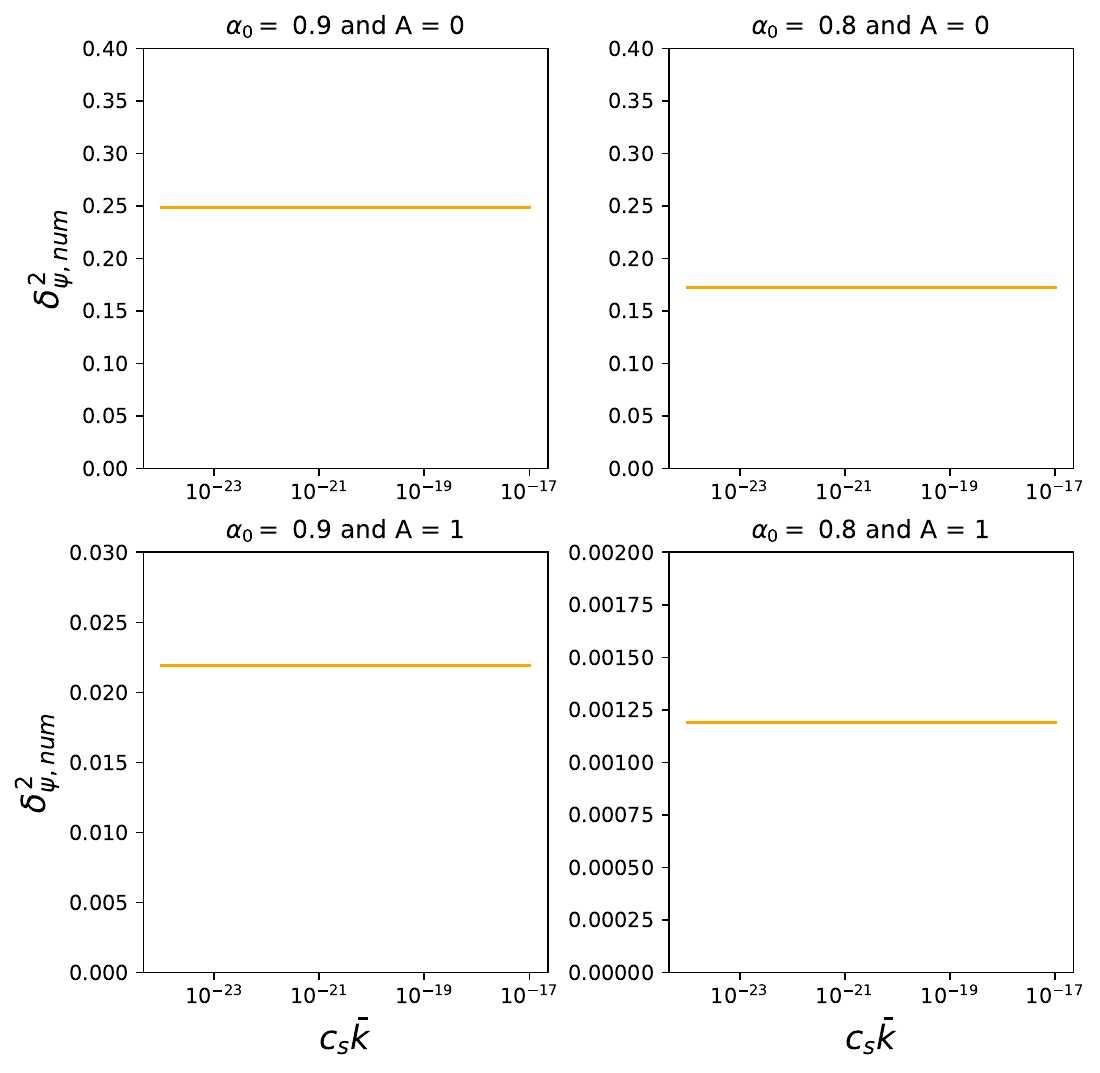}
    \caption{The power spectrum $\delta_{\psi, \rm{num}}^2$  is evaluated for $c_s \bar{k}$ between $10^{-24}$ and $10^{-17}$. The plots show that in the far future, the power spectrum is independent of its value.}
    \label{fig:csk_x_delta_squared}
\end{figure}

\begin{figure*}[htb!]
    \centering
    \includegraphics[width = \textwidth]{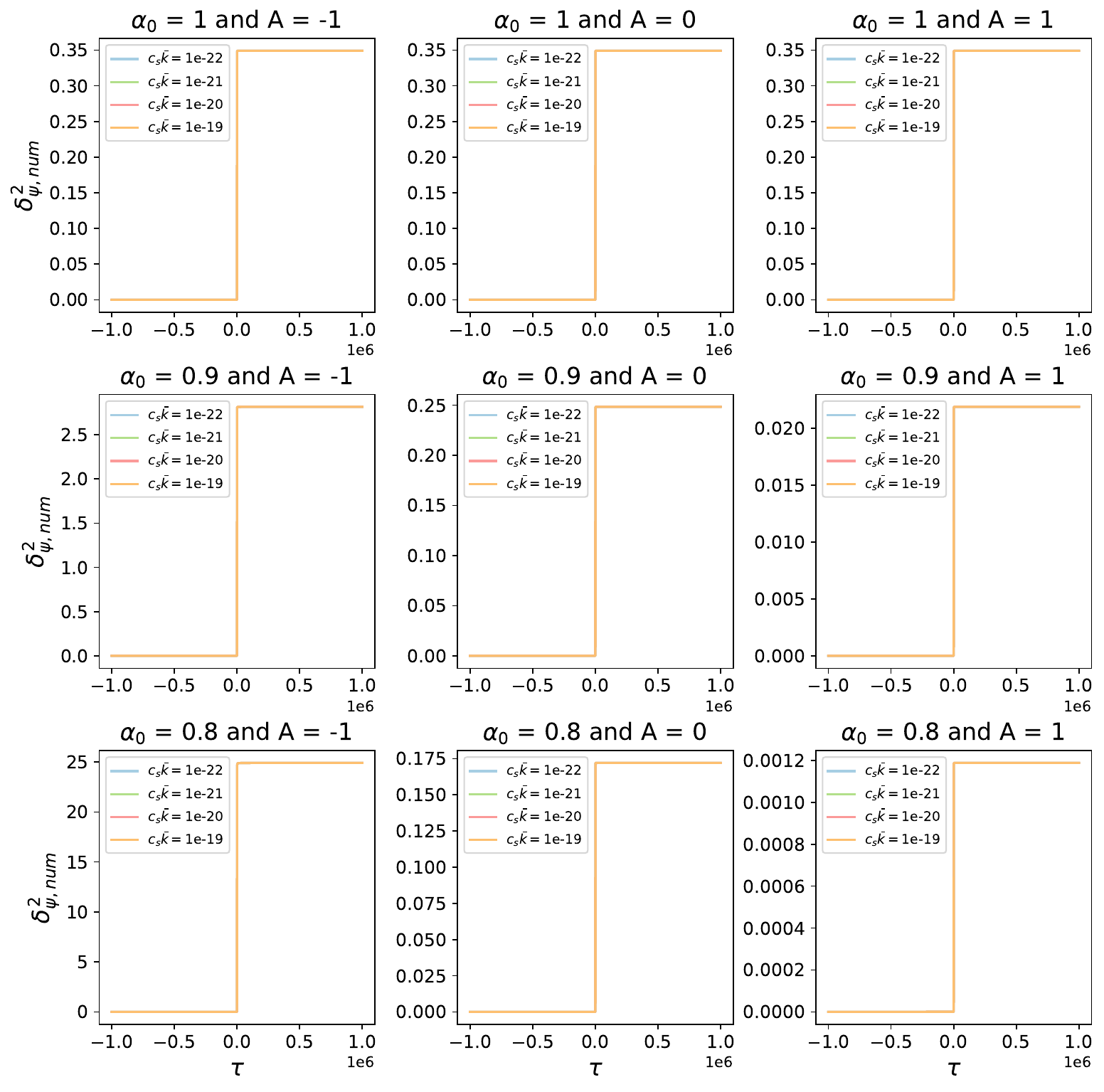}
    \caption{The figures show the time evolution of the power spectrum $\delta_{\psi, \rm{num}}^2$ before and after the bounce, with a huge enhancement after the bounce. All of them have $\tau$ as the horizontal axis, in units of $10^6$. One can see that the power spectrum becomes constant after some time in the expanding phase. The scale invariance is manifest, as the evaluation is made for different $\bar{k} = \{ 10^{-17}, 10^{-16}, 10^{-15}, 10^{-14} \}$, and one cannot distinguish the lines.  The dependence of the amplitude on the free parameters are also exhibited, as the plots are not identical due to the different values appearing on the vertical axes. Note that the amplitude gets larger for negative $A$. The first line presents the $\alpha_0=1$ case, the symmetric bounce, where the power spectrum does not depend on $A$.}
    \label{fig: power_spectrum}
\end{figure*}

The results for $\delta_{\psi, \rm{num}}^2$ are shown in Fig.~\ref{fig:csk_x_delta_squared}, as functions of $\bar{k}$, and in Fig.~\ref{fig: power_spectrum}, as functions of $\tau$. In Fig.~\ref{fig:csk_x_delta_squared}, we take the values of $\delta_{\psi, \rm{num}}^2$ at any value of $\eta$ where it is already constant in time in the expanding phase after the bounce, as it is shown in Fig.~\ref{fig: power_spectrum}. Both figures show the dependence of the amplitude with respect to the free parameters. In Fig.~\ref{fig: power_spectrum} we also show the evolution of $\delta_{\psi, \rm{num}}^2$ for negative $A$, which seems to enlarge its power. One can see in both figures the scale invariance of the spectra for all choices of parameters, as predicted.

From the figures, one can see that, within the relevant background parameters we are taking, $10^{-3}<\delta_{\psi, \rm{num}}^2<10^2$. As Planck data \cite{aghanim2020planck} yield $\delta_{\psi}^2 \approx 10^{-9}$, one can infer from our assumption $c_s \approx \sqrt{\gamma}  =  10^{-5}$ and Eq.~\eqref{power spectrum3} that the possible $t_0$ values are in the range $10^5 l_p < t_0 < 10^9 l_p,$ which are physically reasonable values.

It is worth commenting on the validity of the perturbative treatment, given the low speed of sound $c_s \approx 10^{-5}$. First, there is a quantum strong coupling problem, when third-order corrections of the Lagrangian become bigger than the second-order terms at the first Hubble radius crossing of the perturbations. In the case of inflation, where the amplitude of perturbations is frozen after the first Hubble radius crossing, having the same value as the amplitude of large-scale perturbations we measure today, this constitutes a problem; see Eq. (3.2) of Ref. \cite{Baumann:2011dt}. However, in bouncing cosmologies, the situation is completely different. During the contracting phase, the amplitude of perturbations at the first Hubble radius crossing is much smaller than its value today and continues evolving after it, remaining small over most of the evolution, only growing close to the bounce. Hence, the model does not suffer from the quantum mechanical strong coupling problem, but it may contain classical non-linearities. Nevertheless, as we are considering $w = p/\rho \approx 0$ during contraction, non-Gaussianities turn out to be very small; see Ref. \cite{Baumann:2011dt}, Eqs.~(C2, C8) and Fig.~4 ($\beta=-3$ in this figure). Consequently, the hierarchy between linear and higher-order perturbations is preserved throughout the regime considered here. Also, we assume that eventual curvature corrections to the original Lagrangian, see e.g., Refs. \cite{akrami2020planck, Dutta_2018}, which may cure gradient or ghost instabilities of mimetic gravity, see e.g. Ref. \cite{Ganz_2019, IJJAS2016132}, are not relevant for the present analysis if the bouncing models we obtain do not reach curvature scales where these corrections become relevant.

 \subsection{Comments on stability and the effective nature of the model}

  The higher-derivative mimetic term in Eq.~(\ref{eq: modaction}) gives rise to the effective scalar sound speed \[ c_s^2=\frac{\gamma}{2-3\gamma}. \] Therefore, the parameter range \[ 0<\gamma<\frac{2}{3} \] avoids the short-wavelength gradient instability associated with an imaginary sound speed. Moreover, the mimetic constraint prevents the propagation of an extra higher-derivative degree of freedom so that the possible instability of the scalar sector is not an Ostrogradsky ghost.

  Nevertheless, it is known that the minimal higher-derivative mimetic action may still suffer from an ordinary kinetic ghost once the scalar perturbation action is fully reduced by solving the constraints. In particular, ADM/Hamiltonian analyses of mimetic perturbations have shown that the reduced scalar Hamiltonian is not bounded from below for the minimal action with higher-derivative corrections \cite{Firouzjahi_2017, IJJAS2016132}. This indicates that the minimal model should not be regarded as a complete ultraviolet theory, but rather as a low-energy effective description requiring a UV completion or additional stabilizing operators. Such extensions have been proposed in the literature, which may remove the ghost and gradient instabilities by adding suitable higher-derivative or curvature-dependent operators \cite{Hirano_2017, Gorji_2018}. In the present work, we restrict ourselves to the infrared cosmological regime, relevant for the CMB power spectrum. The verification that a viable UV completion does not affect such infrared calculations is beyond the scope of the present paper.

\subsection{Generalized Potential}

The primordial scalar power spectrum is considered a simple power law, in which its power restricts its slope and is expressed by the spectral index of the initial scalar fluctuations. In particular, when we use the potential ~\eqref{eq: tpotential}, we obtain equations that describe dark matter with zero pressure and a spectral index equal to one, which implies scale invariance. On the other hand, results obtained through cosmic background radiation reveal that the spectral index presents a slight deviation from this value, characterized by a red tilt, where $n_s = 0.965 \pm  0.004$ \cite{akrami2020planck}. Thus, observational data reveal a value that approaches one, but is not exactly one.

To obtain an almost scale invariant $n_s$, we propose a slight modification of the scalar field potential: 
\begin{equation}
        V(\phi) \phi_0^2 = \frac{2(2-3\gamma)}{3} \bigg(\frac{1 + \omega - \omega \frac{\phi^2}{\phi_0^2}}{(1 + \omega)^2 (1 + \frac{\phi^2}{\phi_0^2})^2}\bigg).
         \label{eq: asipotential}
\end{equation}
Following the previous discussion, the scalar field can be identified with time, $\phi = t$, thus $\tau = t/t_0 = \phi/\phi_0$. Substituting these potential into Eq.~\eqref{eq: moddiff}, we find:
\begin{equation}
         \frac{d^2y}{d\tau^2}  - \frac{(1 + \omega - \omega \tau^2)}{(1 + \omega)^2(1+ \tau^2)^2} y(\tau) = 0,
         \label{eq: modpotdiff}
\end{equation}
with solution
\begin{align}
        y(\tau) ={}& C_1 (1+ \tau^2)^{\frac{1}{2(1+\omega)}} + \notag \\
        & 
        C_2(\tau^2 + 1)^{1/(2w + 2)}\tau \times \notag \\
        & 
        \rm{hypergeom}\bigg[\bigg(\frac{1}{2}, \frac{1}{(1 + \omega)}\bigg), \bigg(\frac{3}{2}\bigg), -\tau^2\bigg].
    \label{eq: ynewpothip}
\end{align}
As shown in Subsection~\ref{subsection: Quantization and analytical results}, the adiabatic vacuum condition on sub-Hubble scales selects the mode function \eqref{mode-v vacuum}, which corresponds to setting the integration constant $C_2(k) = 0$. Using this condition, we recover the scale factor solution originally found in Ref.~\cite{bounceNPN1}, namely, \begin{equation}
   a(\tau) = y(\tau)^{2/3} = a_b (1 + \tau^2)^{\frac{1}{3(1 + \omega)}},
 \label{eq: scalefactai}
\end{equation}
yielding the spectral index 
\begin{equation}
   n_s = 1 + \frac{12w}{1+3w}.  
   \label{eq: spectralindex}
\end{equation}
Setting $w<0$ to an appropriate value, we obtain the observed red tilt. The advantage of the mimetic dark matter approach is that $w$ has nothing to do with the sound speed, as in the fluid case, hence setting $w<0$ does not lead to any perturbation instability. Also, as $|w|\ll 1$, the modification will not modify the amplitudes calculated in the previous subsection.


\section{\label{Conclusions} Conclusions}

In this paper, we demonstrated analytically and numerically that mimetic cosmology can yield physically reasonable bouncing models with primordial cosmological perturbations of quantum origin satisfying observational constraints. The model contains only one scalar field, which not only plays the role of dark matter with the appropriate sound velocity but also stops the cosmological contraction at high energies, inducing a classical bounce. To obtain a red-tilted spectral index, one has to make a slight modification to the scalar field potential proposed in Ref.~\cite{chamseddine2014cosmology}, without affecting the sound velocity of the mimetic field perturbations, contrary to what happens with single fluid quantum bouncing models, in which such a slight modification may cause perturbation instabilities \cite{Sandro0}.

The natural length parameter scale of the model $t_0$, which sets the length scale of the bounce, is observationally constrained to stay in the interval $10^5 l_p < t_0 < 10^9 l_p$. Hence, the model can be viewed as a classical bounce scenario induced by the mimetic dark matter field.

Finally, a very important remark: a complete understanding of the ultraviolet completion of the theory and its consequences for the infrared limit is yet to be achieved. At this moment, the calculations we have done indicate a possible way to obtain a bouncing model with a power spectrum of linear scalar perturbations compatible with observations within the mimetic framework. We do not know whether this approach can really be sustained by a healthy theory, but we think that the calculations we have done, the methods we have used, and the satisfactory results we have obtained sofar mighty be useful and applicable to some more fundamental mimetic theory. If this is the case, our results indicate that such theory can be a possible alternative to the primordial origin of the cosmological anisotropies and the structures we see in our universe. Then, the next step would be to use the Markov chain Monte Carlo methods to explore the free parametric space of the model, given by $\alpha_0, A, t_0, w$, obtain the induced cosmological parameters arising from CMB and Supernovae data, and make a Bayesian comparison with inflationary predictions.

\acknowledgments

DRS thanks CNPq for the support under grant 303699/2023-0. ILM expresses his gratitude to CAPES for supporting his PhD project and thanks the Brazilian Center for Research in Physics (CBPF) for its kind hospitality during the development of this work. This study was financed in part by the \textit{Coordenação de Aperfeiçoamento de Pessoal de Nível Superior} (CAPES-Brazil) - Finance Code 001. NPN acknowledges the support of CNPq of Brazil under grant PQ-IB 310121/2021-3. 

\bibliography{ref.bib}  

@article{chamseddine2013mimetic,
  title={Mimetic dark matter},
  author={Chamseddine, Ali H and Mukhanov, Viatcheslav},
  doi = "10.1007/JHEP11%282013%29135",
  journal={Journal of High Energy Physics},
  volume={2013},
  number={11},
  pages={1--5},
  year={2013},
  publisher={Springer}
}

@article{chamseddine2014cosmology,
  title={Cosmology with mimetic matter},
  author={Chamseddine, Ali H and Mukhanov, Viatcheslav and Vikman, Alexander},
  doi = "10.1088/1475-7516/2014/06/017",
  journal={Journal of Cosmology and Astroparticle Physics},
  volume={2014},
  number={06},
  pages={017},
  year={2014},
  publisher={IOP Publishing}
}

@article{vagnozzi2017recovering,
  title={Recovering a MOND-like acceleration law in mimetic gravity},
  author={Vagnozzi, Sunny},
  doi = "10.1088/1361-6382/aa838b",
  journal={Classical and Quantum Gravity},
  volume={34},
  number={18},
  pages={185006},
  year={2017},
  publisher={IOP Publishing}
}

@article{myrzakulov2016static,
  title={Static spherically symmetric solutions in mimetic gravity: rotation curves and wormholes},
  author={Myrzakulov, Ratbay and Sebastiani, Lorenzo and Vagnozzi, Sunny and Zerbini, Sergio},
  doi = "10.1088/0264-9381/33/12/125005",
  journal={Classical and quantum gravity},
  volume={33},
  number={12},
  pages={125005},
  year={2016},
  publisher={IOP Publishing}
}

@article{Baumann:2011dt,
    author = "Baumann, Daniel and Senatore, Leonardo and Zaldarriaga, Matias",
    title = "{Scale-Invariance and the Strong Coupling Problem}",
    eprint = "1101.3320",
    archivePrefix = "arXiv",
    primaryClass = "hep-th",
    doi = "10.1088/1475-7516/2011/05/004",
    journal = "JCAP",
    volume = "05",
    pages = "004",
    year = "2011"
}

@article{bounceNPN1,
    author = "Peter, Patrick and Pinto-Neto, Nelson",
    title = "{Cosmology without inflation}",
    eprint = "0809.2022",
    archivePrefix = "arXiv",
    primaryClass = "gr-qc",
    doi = "10.1103/PhysRevD.78.063506",
    journal = "Phys. Rev. D",
    volume = "78",
    pages = "063506",
    year = "2008"
}

@article{bounceNovello,
    author = "Novello, M. and Bergliaffa, S.E.Perez",
    title = "{Bouncing Cosmologies}",
    eprint = "0802.1634",
    archivePrefix = "arXiv",
    primaryClass = "gr-qc",
    doi = "10.1016/j.physrep.2008.04.006",
    journal = "Phys. Rept.",
    volume = "463",
    pages = "127--213",
    year = "2008"
}

@article{cyclic-Ijjas,
    author = "Ijjas, A. and Steinhardt, P. J.",
    title = "{A new kind of cyclic universe}",
    eprint = "1904.08022",
    archivePrefix = "arXiv",
    primaryClass = "gr-qc",
    doi = "10.1016/j.physletb.2019.06.056",
    journal = "Phys. Lett. B",
    volume = "795",
    pages = "666--672",
    year = "2019"
}

@article{cyclic-Penrose,
    author = "Meissner, A. K. and Penrose, R.",
    title = "{The Physics of Conformal Cyclic Cosmology}",
    eprint = "2503.24263",
    archivePrefix = "arXiv",
    primaryClass = "gr-qc",
    journal = "arXiv preprint arXiv:2503.24263",
    doi = "10.48550/arXiv.2503.24263",
    year = "2025"
}

@article{Ward1,
    author = "Pinto-Neto, N.; Santos, G. B. and Struyve, W.",
    title = "{Quantum-to-classical transition of primordial cosmological perturbations in de Broglie--Bohm quantum theory}",
    eprint = "1110.1339",
    archivePrefix = "arXiv",
    primaryClass = "gr-qc",
    doi = "10.1103/PhysRevD.85.083506",
    journal = "Phys. Rev. D",
    volume = "85",
    pages = "083506",
    year = "2012"
}

@article{Ward2,
    author = "Pinto-Neto, N.; Santos, G. B. and Struyve, W.",
    title = "{Quantum-to-classical transition of primordial cosmological perturbations in de Broglie--Bohm quantum theory: the bouncing scenario}",
    eprint = "1309.2670",
    archivePrefix = "arXiv",
    primaryClass = "gr-qc",
    doi = "10.1103/PhysRevD.89.023517",
    journal = "Phys. Rev. D",
    volume = "89",
    pages = "023517",
    year = "2014"
}

@article{Sandro,
    author = "M. Penna-Lima and N. Pinto-Neto and S. D. P. Vitenti",
    title = "{New formalism to define vacuum states for scalar fields in curved space-times}",
    eprint = "2207.08270",
    archivePrefix = "arXiv",
    primaryClass = "gr-qc",
    doi = "10.1103/PhysRevD.107.065019",
    journal = "Phys. Rev. D",
    volume = "107",
    pages = "065019",
    year = "2023"
}

@misc{DESI,
   author       = {Gu, G. et.al.},
  title        = {Dynamical Dark Energy in light of the DESI DR2 Baryonic Acoustic Oscillations Measurements},
  eprint       = {2504.06118},
  archivePrefix = {arXiv},
  primaryClass = {astro-ph.CO},
  year         = {2025}, 
  note         = {arXiv:2504.06118}
}

@article{Sandro0,
    author = "Vitenti, S. D. P. and Pinto-Neto, N",
    title = "{Adiabatic Scalar Perturbations in a Regular Bouncing Universe}",
    eprint = "1111.0888",
    archivePrefix = "arXiv",
    primaryClass = "gr-qc",
    doi = "10.1103/PhysRevD.85.023524",
    journal = "Phys. Rev. D",
    volume = "85",
    pages = "023524",
    year = "2012"
}

@article{soundspeed,
    author = "Kunz, Martin; Nesseris, Savvas and Sawicki, Ignacy",
    title = "{Constraints on dark-matter properties from large-scale structure}",
    eprint = "1604.05701",
    archivePrefix = "arXiv",
    primaryClass = "astro-ph",
    doi = "10.1103/PhysRevD.94.023510",
    journal = "Phys. Rev. D",
    volume = "94",
    pages = "023510",
    year = "2016"
}

@article{farsi2022structure,
    author = "Farsi, Bita and Sheykhi, Ahmad",
    title = "{Structure formation in mimetic gravity}",
    eprint = "2202.04118 ",
    archivePrefix = "arXiv",
    primaryClass = "gr-qc",
    doi = "10.1103/PhysRevD.106.024053",
    journal = "Phys. Rev. D",
    volume = "106",
    pages = "024053",
    year = "2022"
}

@article{chamseddine2017resolving,
    author = "Chamseddine, Ali H. and Mukhanov, Viatcheslav",
    title = "{Resolving cosmological singularities}",
    eprint = "1612.05860",
    archivePrefix = "arXiv",
    primaryClass = "gr-qc",
    doi = "10.1088/1475-7516/2017/03/009",
    journal = "J. Cosmol. Astropart. Phys.",
    volume = "03",
    number = "03",
    pages = "009",
    year = "2017"
}

@article{chamseddine2017nonsingular,
    author = "Chamseddine, Ali H and Mukhanov, Viatcheslav",
    title = "{Nonsingular black hole}",
    eprint = "1612.05861 ",
    archivePrefix = "arXiv",
    primaryClass = "gr-qc",
    doi = "10.1140/epjc/s10052-017-4759-z",
    journal = "Eur. Phys. J. C",
    volume = "77",
    pages = "183",
    year = "2017"
}

@article{sheykhi2021topological,
    author = "Sheykhi, A. and Grunau, S.",
    title = "{Topological black holes in mimetic gravity}",
    eprint = "1911.13072",
    archivePrefix = "arXiv",
    primaryClass = "gr-qc",
    doi = "10.1142/S0217751X21501864",
    journal = "International Journal of Modern Physics A",
    volume = "36",
    pages = "1--27",
    year = "2021"
}

@article{chen2018black,
    author = "Chen, C. Y. and Bouhmadi-López, M.  and Chen, P.",
    title = "{Black hole solutions in mimetic Born-Infeld gravity}",
    eprint = "1710.10638",
    archivePrefix = "arXiv",
    primaryClass = "gr-qc",
    doi = "10.1140/epjc/s10052-018-5556-z",
    journal = "Eur. Phys. J. C",
    volume = "78",
    pages = "1--9",
    year = "2018"
}

@article{nashed2022black,
    author = "Nasheda, G.G.L. and Shin’ichi Nojiri",
    title = "{Black holes with Lagrange multiplier and potential in mimetic-like gravitational theory: multi-horizon black holes}",
    eprint = "2110.08560",
    archivePrefix = "arXiv",
    primaryClass = "gr-qc",
    doi = "10.1088/1475-7516/2022/05/011",
    journal = "J. Cosmol. Astropart. Phys.",
    volume = "2022",
    number = "05",
    pages = "011",
    year = "2022"
}

@article{casalino2018mimicking,
    author = "Casalino, A. and Rinaldi, M. and Sebastiani, L. and Vagnozzi, S.",
    title = "{Mimicking dark matter and dark energy in a mimetic model compatible with GW170817}",
    eprint = "1803.02620",
    archivePrefix = "arXiv",
    primaryClass = "gr-qc",
    doi = "10.1016/j.dark.2018.10.001",
    journal = "Phys. Dark Univ.",
    volume = "22",
    pages = "108--115",
    year = "2018"
}

@article{Casalino2019alive,
    author = "Casalino, A. and Massimiliano, R. and Sebastiani, L. and Vagnozzi, S.",
    title = "{Alive and well: mimetic gravity and a higher-order
extension in light of GW170817}",
    eprint = "1811.06830",
    archivePrefix = "arXiv",
    primaryClass = "gr-qc",
    doi = "10.1088/1361-6382/aaf1fd",
    journal = "Class. Quantum Grav.",
    volume = "36",
    pages = "1--15",
    year = "2019"
}

@article{Sharafati2021higher,
    author = "Sharafati, K. and Heydari, S. and Karami, K.",
    title = "{Higher-Order mimetic gravity after GW170817}",
    eprint = "2109.11810",
    archivePrefix = "arXiv",
    primaryClass = "gr-qc",
    doi = "10.1142/S0217732323500207",
    journal = "Modern Physics Letters A",
    volume = "38",
    pages = "2350020",
    year = "2023"
}

@article{Dutta_2018,
doi = {10.1088/1475-7516/2018/02/041},
url = {https://doi.org/10.1088/1475-7516/2018/02/041},
year = {2018},
month = {feb},
publisher = {},
volume = {2018},
number = {02},
pages = {041},
author = {Dutta, Jibitesh and Khyllep, Wompherdeiki and Saridakis, Emmanuel N. and Tamanini, Nicola and Vagnozzi, Sunny},
title = {Cosmological dynamics of mimetic gravity},
journal = {Journal of Cosmology and Astroparticle Physics}
}

@article{IJJAS2016132,
title = {NEC violation in mimetic cosmology revisited},
journal = {Physics Letters B},
volume = {760},
pages = {132-138},
year = {2016},
issn = {0370-2693},
doi = {https://doi.org/10.1016/j.physletb.2016.06.052},
url = {https://www.sciencedirect.com/science/article/pii/S0370269316303124},
author = {Anna Ijjas and Justin Ripley and Paul J. Steinhardt}
}

@article{Firouzjahi_2017,
doi = {10.1088/1475-7516/2017/07/031},
url = {https://doi.org/10.1088/1475-7516/2017/07/031},
year = {2017},
month = {jul},
publisher = {},
volume = {2017},
number = {07},
pages = {031},
author = {Firouzjahi, Hassan and Gorji, Mohammad Ali and Mansoori, Seyed Ali Hosseini},
title = {Instabilities in mimetic matter perturbations},
journal = {Journal of Cosmology and Astroparticle Physics}
}

@article{Hirano_2017,
doi = {10.1088/1475-7516/2017/07/009},
url = {https://doi.org/10.1088/1475-7516/2017/07/009},
year = {2017},
month = {jul},
publisher = {},
volume = {2017},
number = {07},
pages = {009},
author = {Hirano, Shin'ichi and Nishi, Sakine and Kobayashi, Tsutomu},
title = {Healthy imperfect dark matter from effective theory of mimetic cosmological perturbations},
journal = {Journal of Cosmology and Astroparticle Physics}
}

@article{Gorji_2018,
doi = {10.1088/1475-7516/2018/01/020},
url = {https://doi.org/10.1088/1475-7516/2018/01/020},
year = {2018},
month = {jan},
publisher = {},
volume = {2018},
number = {01},
pages = {020},
author = {Gorji, Mohammad Ali and Mansoori, Seyed Ali Hosseini and Firouzjahi, Hassan},
title = {Higher derivative mimetic gravity},
journal = {Journal of Cosmology and Astroparticle Physics}
}

@article{Ganz_2019,
doi = {10.1088/1475-7516/2019/12/037},
url = {https://doi.org/10.1088/1475-7516/2019/12/037},
year = {2019},
month = {dec},
publisher = {},
volume = {2019},
number = {12},
pages = {037},
author = {Ganz, Alexander and Bartolo, Nicola and Matarrese, Sabino},
title = {Towards a viable effective field theory of mimetic gravity},
journal = {Journal of Cosmology and Astroparticle Physics}
}

@article{Chen2021thick,
    author = "Jing Chen and Wen-Di Guo and Yu-Xiao Liu",
    title = "{Thick branes with inner structure in mimetic f(R) gravity}",
    eprint = "2011.03927",
    archivePrefix = "arXiv",
    primaryClass = "gr-qc",
    doi = "10.48550/arXiv.2011.03927",
    journal = "Eur. Phys. J. C",
    volume = "81",
    number = "709",
    pages = "1--16",
    year = "2021"
}

@article{Mansoori2022multi,
    author = "Mansoori, Seyed Ali Hosseini et al.",
    title = "{Multifield mimetic gravity}",
    eprint = "2108.11666",
    archivePrefix = "arXiv",
    primaryClass = "gr-qc",
    doi = "10.1103/PhysRevD.105.023529",
    journal = "Phys. Rev. D",
    volume = "105",
    number = "2",
    pages = "023529",
    year = "2022"
}

@article{ramo2019cosmological,
    author = "Ramo Chothe, H. and Dutta, A. and Sur, S.",
    title = "{Cosmological dark sector from a mimetic–metric–torsion perspective}",
    eprint = "1907.12429",
    archivePrefix = "arXiv",
    primaryClass = "gr-qc",
    doi = "10.1142/S02182718195017488",
    journal = "International Journal of Modern Physics D",
    volume = "28",
    number = "15",
    pages = "1950174",
    year = "2019",
    publisher={World Scientific}
}

@article{kaczmarek2024mimetic,
      title={Mimetic-$f(Q)$ gravity: Cosmic reconstruction and energy conditions}, 
      author={Kaczmarek, Adam Z},
      doi = { 10.1016/j.nuclphysb.2024.116677},
      journal={Nuclear Physics B},
      volume={1007},
      pages={116677},
      year={2024},
      publisher={Elsevier},
      archivePrefix = "arXiv",
      primaryClass = "gr-qc",
      eprint ={2401.04084}, 
}

@article{ijjas2018bouncing,
  author    = {Ijjas, Anna and Steinhardt, Paul J},
  title     = {Bouncing Cosmology Made Simple},
  journal   = {Classical and Quantum Gravity},
  volume    = {35},
  number    = {13},
  pages     = {135004},
  year      = {2018},
  publisher = {IOP Publishing},
  doi       = {10.1088/1361-6382/aac482},
}

@article{aghanim2020planck,
  author    = {Planck Collaboration and N. Aghanim and Y. Akrami and M. Ashdown and J. Aumont and C. Baccigalupi and et al.},
  title     = {Planck 2018 results. VI. Cosmological parameters},
  journal   = {A\&A},
  volume    = {641},
  year      = {2020},
  pages     = {A6},
  doi       = {10.1051/0004-6361/201833910},
  archivePrefix = "arXiv",
  primaryClass = "astro-ph.CO",
  eprint     = {1807.06209},
}

@article{akrami2020planck,
  title={Planck 2018 results-X. Constraints on inflation},
  author={Akrami, Yashar and Arroja, Frederico and Ashdown, M and Aumont, J and Baccigalupi, Carlo and Ballardini, M and Banday, Anthony J and Barreiro, RB and Bartolo, Nicola and Basak, S and others},
  journal={Astronomy \& Astrophysics},
  volume={641},
  pages={A10},
  year={2020},
  publisher={EDP sciences},
  doi = {10.1051/0004-6361/201833887},
  archivePrefix = "arXiv",
  primaryClass = "astro-ph.CO",
}

@book{Mukhanov2005BookOfCosmology,
  title={Physical foundations of cosmology},
  author={Mukhanov, Viatcheslav},
  year={2005},
  publisher={Cambridge University Press}
}

@article{scipy,
  title={SciPy 1.0: fundamental algorithms for scientific computing in Python},
  author={Virtanen, Pauli and Gommers, Ralf and Oliphant, Travis E and Haberland, Matt and Reddy, Tyler and Cournapeau, David and Burovski, Evgeni and Peterson, Pearu and Weckesser, Warren and Bright, Jonathan and others},
  journal={Nature methods},
  doi = "10.1038/s41592-019-0686-2",
  volume={17},
  number={3},
  pages={261--272},
  year={2020},
  publisher={Nature Publishing Group US New York}
}

@article{matsumoto2015cosmological,
  title={Cosmological perturbations in a mimetic matter model},
  author={Matsumoto, Jiro and Odintsov, Sergei D and Sushkov, Sergey V},
  journal={Physical Review D},
  doi = "https://doi.org/10.1103/PhysRevD.91.064062",
  volume={91},
  number={6},
  pages={064062},
  year={2015},
  publisher={APS}
}

@article{nojiri2014mimetic,
  title={Mimetic F(R) gravity: inflation, dark energy and bounce},
  author={Nojiri, Shin'ichi and Odintsov, Sergei D},
  journal={Modern Physics Letters A},
  doi = "https://doi.org/10.1142/S0217732314502113",
  volume={29},
  number={40},
  pages={1450211},
  year={2014},
  publisher={World Scientific}
}

@article{calza2024implications,
  title={Implications of cosmologically coupled black holes for pulsar timing arrays},
  author={Calz{\`a}, Marco and Gianesello, Francesco and Rinaldi, Massimiliano and Vagnozzi, Sunny},
  DOI = "https://doi.org/10.1038/s41598-024-82661-8",
  journal={Scientific Reports},
  volume={14},
  number={1},
  pages={31296},
  year={2024},
  publisher={Nature Publishing Group UK London}
}

@article{myrzakulov2015inflation,
  title={Inflation in f (R, $\phi$) f (R, $\phi$)-theories and mimetic gravity scenario},
  author={Myrzakulov, Ratbay and Sebastiani, Lorenzo and Vagnozzi, Sunny},
  doi = "https://doi.org/10.1140/epjc/s10052-015-3672-6",
  journal={The European Physical Journal C},
  volume={75},
  pages={1--11},
  year={2015},
  publisher={Springer}
}

@article{cognola2016covariant,
  title={Covariant Ho{\v{r}}ava-like and mimetic Horndeski gravity: cosmological solutions and perturbations},
  author={Cognola, Guido and Myrzakulov, Ratbay and Sebastiani, Lorenzo and Vagnozzi, Sunny and Zerbini, Sergio},
  doi = "https://doi.org/10.1088/0264-9381/33/22/225014",
  journal={Classical and quantum gravity},
  volume={33},
  number={22},
  pages={225014},
  year={2016},
  publisher={IOP Publishing}
}

@article{jirouvsek2022disforming,
  title={Disforming to conformal symmetry},
  author={Jirou{\v{s}}ek, Pavel and Shimada, Keigo and Vikman, Alexander and Yamaguchi, Masahide},
  journal={Journal of Cosmology and Astroparticle Physics},
  doi = "https://doi.org/10.1088/1475-7516/2022/11/019",
  volume={2022},
  number={11},
  pages={019},
  year={2022},
  publisher={IOP Publishing}
}

@article{lim2010dust,
  title={Dust of dark energy},
  author={Lim, Eugene A and Sawicki, Ignacy and Vikman, Alexander},
  doi ="https://doi.org/10.1088/1475-7516/2010/05/012",
  journal={Journal of Cosmology and Astroparticle Physics},
  volume={2010},
  number={05},
  pages={012},
  year={2010},
  publisher={IOP Publishing}
}

@article{capozziello2010dark,
  title={Dark energy from modified gravity with Lagrange multipliers},
  author={Capozziello, Salvatore and Matsumoto, Jiro and Nojiri, Shin'ichi and Odintsov, Sergei D},
  doi = "https://doi.org/10.1016/j.physletb.2010.08.030",
  journal={Physics Letters B},
  volume={693},
  number={2},
  pages={198--208},
  year={2010},
  publisher={Elsevier}
}

@article{gao2011cosmological,
  title={Cosmological models with Lagrange multiplier field},
  author={Gao, Changjun and Gong, Yan and Wang, Xin and Chen, Xuelei},
  journal={Physics Letters B},
  doi = "https://doi.org/10.1016/j.physletb.2011.06.085",
  volume={702},
  number={2-3},
  pages={107--113},
  year={2011},
  publisher={Elsevier}
}

@article{sebastiani2017mimetic,
  title={Mimetic gravity: a review of recent developments and applications to cosmology and astrophysics},
  author={Sebastiani, Lorenzo and Vagnozzi, Sunny and Myrzakulov, Ratbay},
  doi = "https://doi.org/10.1155/2017/3156915",
  journal={Advances in High Energy Physics},
  volume={2017},
  number={1},
  pages={3156915},
  year={2017},
  publisher={Wiley Online Library}
}

@article{capela2015modified,
  title={Modified dust and the small scale crisis in CDM},
  author={Capela, Fabio and Ramazanov, Sabir},
  doi = "https://doi.org/10.1088/1475-7516/2015/04/051",
  journal ={Journal of Cosmology and Astroparticle Physics},
  volume={2015},
  number={04},
  pages={051},
  year={2015},
  publisher={IOP Publishing}
}

@article{ramazanov2016living,
  title={Living with ghosts in Ho{\v{r}}ava-Lifshitz gravity},
  author={Ramazanov, Sabir and Arroja, F and Celoria, M and Matarrese, Sabino and Pilo, L},
  doi = "https://doi.org/10.1007/JHEP06%282016%29020",
  journal ={Journal of High Energy Physics},
  volume={2016},
  number={6},
  pages={1--31},
  year={2016},
  publisher={Springer}
}

@article{babichev2017gravitational,
  title={Gravitational focusing of imperfect dark matter},
  author={Babichev, Eugeny and Ramazanov, Sabir},
  doi = "https://doi.org/10.1103/PhysRevD.95.024025",
  journal={Physical Review D},
  volume={95},
  number={2},
  pages={024025},
  year={2017},
  publisher={APS}
}

\appendix

\end{document}